\documentclass[11pt,a4paper]{article}
\pdfoutput=1
\usepackage{jcappub}
\usepackage[english]{babel} 					
\usepackage{fancyhdr} 						
\usepackage{times} 						
\usepackage{amsmath} 						
\usepackage{amsfonts} 						
\usepackage{amssymb} 						
\usepackage{ifthen}   						
\usepackage{textcomp}						
\usepackage{pifont}						
\usepackage{units}						
\usepackage{tabularx}
\usepackage{esint}
\usepackage{multicol}
\usepackage{lscape}
\usepackage{url}
\usepackage{gensymb}
\usepackage{rotating,capt-of}
\usepackage{multirow}
\usepackage{wrapfig}
\usepackage{float}
\usepackage{natbib}
\usepackage{placeins}
\usepackage{booktabs}

\title{A sensitive search for unknown spectral emission lines in the diffuse X-ray background with XMM-Newton}
\author[a]{A. Gewering-Peine}
\author[a]{D. Horns}
\author[b]{J.H.M.M. Schmitt}

\affiliation[a]{Institut f\"ur Experimentalphysik, Universit\"at Hamburg,\\
Luruper Chaussee 149, D-22761 Hamburg, Germany}
\affiliation[b]{Hamburger Sternwarte, Universit\"at Hamburg,\\
Gojenbergsweg 112, D-21029 Hamburg, Germany}

\emailAdd{alexander.gewering-peine@desy.de}
\emailAdd{dieter.horns@desy.de}
\emailAdd{jschmitt@hs.uni-hamburg.de}

\abstract{The Standard Model of particle physics can be extended to include sterile (right-handed) neutrinos or axions to solve the dark matter problem. Depending upon the mixing angle between active and sterile neutrinos, the latter have the possibility to decay into monoenergetic active neutrinos and photons in the keV-range while axions can couple to two photons. We have used data taken with X-ray telescopes like e.g. XMM-Newton for the search of line emissions. We used pointings with high exposures and expected dark matter column densities with respect to the dark matter halo of the Milky Way. The posterior predictive p-value analysis has been applied to locate parameter space regions which favour additional emission lines. In addition, upper limits of the parameter space of the models have been generated such that the preexisting limits have been significantly improved.}


\begin{document}
\maketitle

\begin{abstract}

\end{abstract}

\section{Introduction}

The $\Lambda$CDM\footnote{The term $\Lambda$CDM is an abbreviation composed of the greek letter $\Lambda$ which stands for a cosmological constant and CDM which stands for "Cold Dark Matter``.} model describes the universe on cosmological scales and requires a non-baryonic dark matter component \cite{Ade:2015xua}. The rotation curves of galaxies including the Milky Way \cite{Sofue:2000jx,Iocco:2015xga} reveal an unexpected flatness which can be explained by the gravitational influence of such a dark matter component. The dark matter is distributed as a spherical halo. The shape of the dark matter halo derived from the observations allows to interpret its constituents as non-relativistic and non-baryonic particles. The simplest approach to describe the dark matter halo is an isothermal sphere.\\
Most of the dark matter particle models predict weak interactions of the dark matter with baryonic matter which opens an observation window to the dark matter component. These dark matter particle candidates fall into three groups with widely different rest masses: (1) weakly interacting massive particles (WIMPs), (2) ultra-light and cold weakly interacting slim particles (WISPs) and (3) warm dark matter particles with energies in the keV to MeV range like sterile neutrinos.
This work investigates the predicted observable effects related to the particles of categories (2) and (3) by the search for unidentified emission lines in the spectra of the diffuse X-ray background which could originate in the decay of sterile neutrinos or the transition of axions into photons. Such a signal is expected if the particles with mass $m=\mathcal{O}(\mathrm{keV})$ are unstable on cosmological time-scales and decay into a two-particle final-state.\\
Previous publications have dealt with the observation of three major object classes to derive upper limits of parameter spaces of dark matter models: spiral galaxies \cite{Anderson:2014tza,Borriello:2011un,Watson:2006qb}, especially the Milky Way \cite{Jeltema:2014qfa,Malyshev:2014xqa,Boyarsky:2014ska,Riemer-Sorensen:2014yda,Boyarsky:2006ag}, and M31 \cite{Boyarsky:2014paa,Jeltema:2014qfa,Boyarsky:2014ska,Boyarsky:2014jta,Horiuchi:2013noa,Watson:2011dw,Boyarsky:2007ay,Watson:2006qb}, dwarf spheroidals \cite{Malyshev:2014xqa,Boyarsky:2006ag}, galaxy clusters \cite{Bulbul:2014sua,Boyarsky:2014jta,Boyarsky:2006kc}, and the diffuse galactic and extragalactic background \cite{Boyarsky:2014jta,Essig:2013goa,Boyarsky:2005us}. The data of the X-ray telescopes XMM-Newton, Chandra and HEAO-1 have been primarily used for the analyses in these publications. The typical exposures and expected dark matter column densities lead to an order of $\mathcal{O}(10-1000)\,\text{ks}$ and $\mathcal{O}(10^{28})\,\text{keV cm$^{-2}$}$, respectively. The detection of a unknown spectral emission line at an energy of $3.55\,$keV was claimed \cite{Bulbul:2014sua,Boyarsky:2014jta} and gave rise to intensive discussions and further analyses \cite{Boyarsky:2014paa,Jeltema:2014qfa,Malyshev:2014xqa,Anderson:2014tza,Boyarsky:2014ska,Riemer-Sorensen:2014yda,Franse:2016dln,Aharonian:2016gzq,Shah:2016efh} as well as several theoretical interpretations \cite{Ishida:2014dlp,Abazajian:2014gza,Baek:2014qwa,Tsuyuki:2014aia,Allahverdi:2014dqa,Modak:2014vva,Cline:2014eaa,Robinson:2014bma,Abada:2014zra,Okada:2014oda,Haba:2014taa,Finkbeiner:2014sja,Okada:2014zea,Higaki:2014zua,Jaeckel:2014qea,Lee:2014xua,Cicoli:2014bfa,Kong:2014gea,Choi:2014tva,Dias:2014osa,Liew:2014gia,Conlon:2014xsa,Bomark:2014yja,Demidov:2014hka,Nakayama:2014ova,Chiang:2014xra,Kolda:2014ppa,Dutta:2014saa,Queiroz:2014yna,Lee:2014koa,Geng:2014zqa,Cline:2014kaa,Krall:2014dba,Frandsen:2014lfa,Dudas:2014ixa,Baek:2014poa}. A consensus on the origin of this line has not been arrived at so far. Future X-ray missions as eRosita \cite{Zandanel:2015xca} and Athena \cite{Neronov:2015kca} will likely have the chance to investigate this topic. Unfortunately, the Astro-E mission failed due to a technical failure of the spacecraft.\\
In the aforementioned publications \cite{Boyarsky:2014paa,Jeltema:2014qfa,Malyshev:2014xqa,Anderson:2014tza,Boyarsky:2014ska,Riemer-Sorensen:2014yda,Franse:2016dln} widely different statistical methods have been used which may not be optimal in coverage and sensitivity. This work presents an optimised data selection and statistical method. All available data sets of the X-ray telescope XMM-Newton \cite{2012arXiv1202.1651L} have been ranked by the exposure and the predicted dark matter column densities. The diffuse X-ray background spectra of the most promising data sets according to the ranking have been generated. The posterior predictive p-value analysis \cite{Protassov:2002sz} was chosen as the statistical method to determine the upper limits of possible unknown Gaussian-shaped emission lines in the background spectra mentioned before. This method works as a hypothesis test among spectral models of different degrees of freedom, e.g., an continuum model and the same model plus an additive component (Gaussian emission line). This statistical method has the advantage that the underlying sampling distributions, generated by Monte-Carlo processes, are well-known, which is not true for the pure application of the F-test or the likelihood-ratio method.\\

\section{Observations and Data Analysis}

\subsection{Data selection}

This work uses data sets up to the year $2013$ taken with the EPIC PN detector \cite{Struder:2001bh} onboard the XMM-Newton satellite since it is the X-ray detector with the largest effective area available. Two features are directly relevant to the filtering of the most promising data sets:
\begin{enumerate}
 \item The exposure time should be maximised to reach high signal-to-noise ratios (S/N) to derive the most sensitive limits of emission lines in general.
 \item The direction of the pointed observations is directly linked to the model-dependent column density of the dark matter distribution.
\end{enumerate}
A correspondingly meaningful benchmark value for each data set was constructed under assumption of an Navarro-Frenk-White (NFW) dark matter profile \cite{Navarro:1995iw} with the shape
\begin{align}
   \rho_{\,\text{NFW}}(r)=\frac{\rho_0}{\left(r/r_0\right)\left(1+r/r_0\right)^2}\quad\text{and}\quad\rho_0=0.4\,\frac{\text{GeV}}{c^2}\,\text{cm}^{-3},\,r_0=21\,\text{kpc}.                                    
\end{align}
The normalised product of the $i$th of $N$ data sets obtained from the estimated or raw exposure $t_{\,\text{exp};i}$ and the dark matter column density $S_{\,\text{NFW};i}$, dependent on the integration length $s$ and the Galactic coordinates $(l,b)$, followed by equation \ref{SDM}, was chosen as the benchmark value:
\begin{align}
 b_{\,\text{NFW};i}=\frac{S_{\,\text{NFW};i}\cdot t_{\,\text{exp};i}}{\max\left(S_{\text{NFW};1}\cdot t_{\,\text{exp};1},\ldots,S_{\text{NFW};j}\cdot t_{\,\text{exp};j},\ldots,S_{\text{NFW};N}\cdot t_{\,\text{exp};N}\right)},\quad i\in\left\{1,\ldots,N\right\},
\end{align}
where
\begin{align}
 S_{\,\text{NFW}}(s,l_i,b_i)=S_{\,\text{NFW};i}=\int_0^{\infty}\,\rho_{\text{dm}}(s,l_i,b_i)\,ds, \label{SDM}.
\end{align}

The highly time-varying instrumental and particle backgrounds encountered by XMM-Newton are not covered by this benchmark. Despite this disadvantage, the benchmark is acceptable because the most common dark matter distribution models have a comparable spatial and angular shape and a density maximum matching the position of the Galactic center. In this work, the focus is set onto the presumed dark matter halo of our Milky Way. Table \ref{tab:datasets} presents the specifications of the data sets. The observation identifications (obsid) as well as the applied filters\footnote{The 'Thin1' and 'Medium' filters \cite{Struder:2001bh} consist of $40$\,nm and $80$\,nm aluminium plus $160$\,mn polymide, respectively.} are listed. Furthermore, the Galactic coordinates ($l$,$b$), the estimated and the exposures after filtering (net exposures) as well as the net fields of view (net fov) are shown next to the expected dark matter column densities $S_{\text{NFW}}$ based on the NFW-distribution model. The very right column contains the final benchmark values $b_{\text{NFW},i}$.
The poor quality of a major part of the data sets in terms of a high contamination of instrumental and particle background by cosmic-ray induced flourescence processes and/or soft proton clouds collected by the detectors (see section $2.2$), respectively, led to a tight reduction of the number of data sets used for analysis. In practice, the data sets with the highest benchmark values and net exposures above $22\,\text{ks}$, listed in table \ref{tab:datasets}, have been used for further analysis.

\begin{table}[ht]
\centering\small\setlength\tabcolsep{5pt}
\renewcommand\arraystretch{1.3}
\scalebox{0.875}{
\begin{tabular}{cccccccccccc}
\hline
Object & Obsid	  	&Filter	&$l$			&$b$			&Estimated 	&Net		& Net &$S_{\,\text{NFW}}$ &$b_{\,\text{NFW}}$\\
& &		&	&					&exposure	&exposure	& fov 	&						& \\
\hline
& &	&	degree	&	degree	& 	s	&	s	& $10^{-5}\,\text{str}$ &  $10^{27}\,\frac{\text{keV}}{\text{cm}^{2}}$&	 \\ \hline\hline
V4046 Sgr&0604860201&Medium&359.66&-7.26&123316.0&62580.4&13.45&164.38&0.86\\ 
V4046 Sgr&0604860301&Medium&359.66&-7.26&123419.0&72761.8&13.78&164.38&1.00\\ 
V4046 Sgr&0604860401&Medium&359.66&-7.26&123715.0&72572.7&13.63&164.38&1.00\\ 
V4633 Sgr&0653550301&Thin1&5.12&-6.26&87320.0&56167.0&13.71&158.87&0.75\\ 
PDS 456&0501580201&Thin1&10.38&11.14&89711.0&62158.3&13.31&125.92&0.65\\ 
PDS 456&0501580101&Thin1&10.38&11.14&92359.0&49372.2&12.46&125.92&0.52\\ 
VV Sco&0555650301&Medium&352.63&19.86&105375.0&40931.0&13.60&108.61&0.37\\ 
VV Sco&0555650201&Medium&352.63&19.86&104418.0&48848.0&13.61&108.61&0.44\\ 
CNOC2 Field 1&0603590101&Medium&50.98&-42.00&82317.0&34084.2&13.58&54.42&0.16\\ 
OGLE 1999 BUL 32&0152420101&Medium&2.45&-3.53&49940.0&28501.1&12.22&191.69&0.46\\ 
MACHO 96 BLG 5&0305970101&Thin1&3.21&-3.10&107012.0&48967.2&12.93&189.78&0.78\\ 
LH VLA 2&0554121301&Medium&148.46&51.42&55542.0&22947.6&13.56&30.45&0.06\\ 
RXJ2328.8+1453&0502430301&Thin1&94.96&-43.46&104910.0&53772.2&13.53&38.07&0.17\\ 
CDFS&0555780101&Thin1&223.46&-54.40&133118.0&34855.0&13.44&31.92&0.09\\ 
CDFS&0555780201&Thin1&223.48&-54.40&133416.0&42680.9&13.47&31.92&0.11\\ 
CDFS&0555780301&Thin1&223.47&-54.41&123811.0&50978.4&13.54&31.92&0.14\\ 
CDFS&0555780501&Thin1&223.63&-54.43&113004.0&61887.7&13.62&31.94&0.17\\ 
CDFS&0555780601&Thin1&223.64&-54.43&118413.0&48940.7&13.68&31.94&0.13\\ 
CDFS&0555780701&Thin1&223.66&-54.43&118415.0&59393.9&13.59&31.94&0.16\\ 
CDFS&0555780801&Thin1&223.62&-54.44&120919.0&47766.2&13.50&31.94&0.13\\ 
CDFS&0555780901&Thin1&223.64&-54.44&121518.0&44934.8&13.52&31.94&0.12\\ 
CDFS&0555781001&Thin1&223.65&-54.44&125813.0&58772.7&13.65&31.94&0.16\\ 
CDFS&0555782301&Thin1&223.65&-54.44&125714.0&51584.6&13.64&31.94&0.14\\ 

\hline
\end{tabular}
}
\caption{\label{tab:datasets} The table lists the observation IDs, coordinates and the estimated and net exposures as well as the net fields of view of the data sets analysed in this publication. The model-dependent, predicted dark matter column densities $S_{\text{DM}}$ and the final benchmark values $b_{\text{NFW}}$ are also listed.}
\end{table}

\subsection{SAS Data Reduction and Analysis}

\begin{figure}
\includegraphics[width=1.0\textwidth]{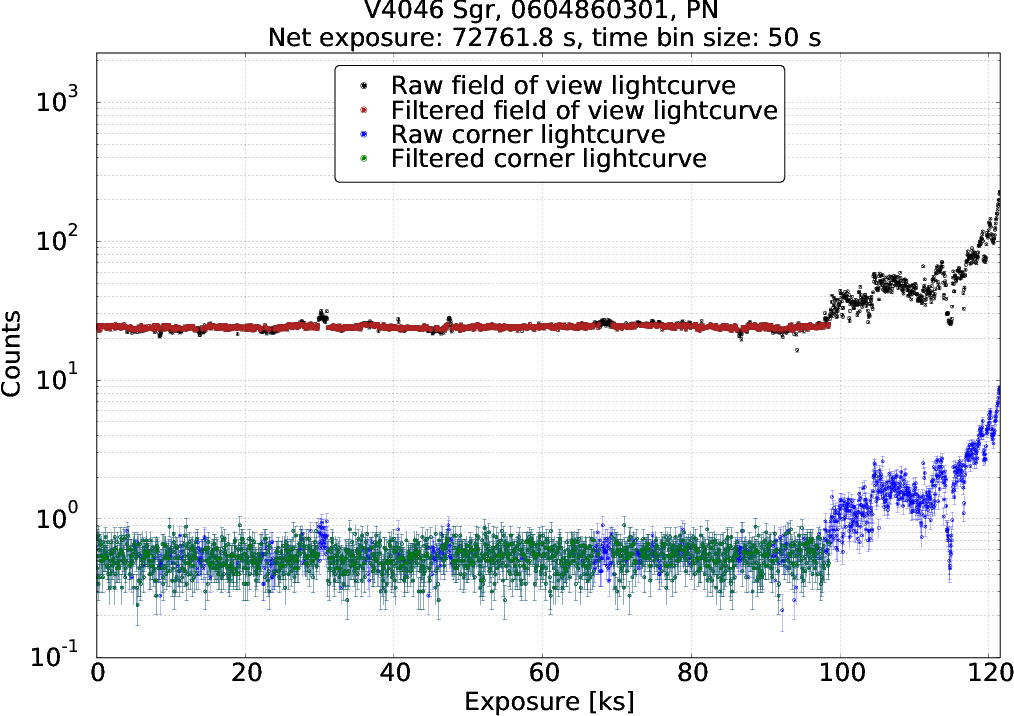}
\caption{Lightcurves of the field of view and of the corners of the chip as well as the filtered lightcurves of the data set $0604860301$, selected for the successive analysis.}
\label{fig:TOP-TEN-0604860301-PN-4-bkg-380-12000-lightcurve-files26112015}
\end{figure}

\begin{figure}
\includegraphics[width=1.0\textwidth]{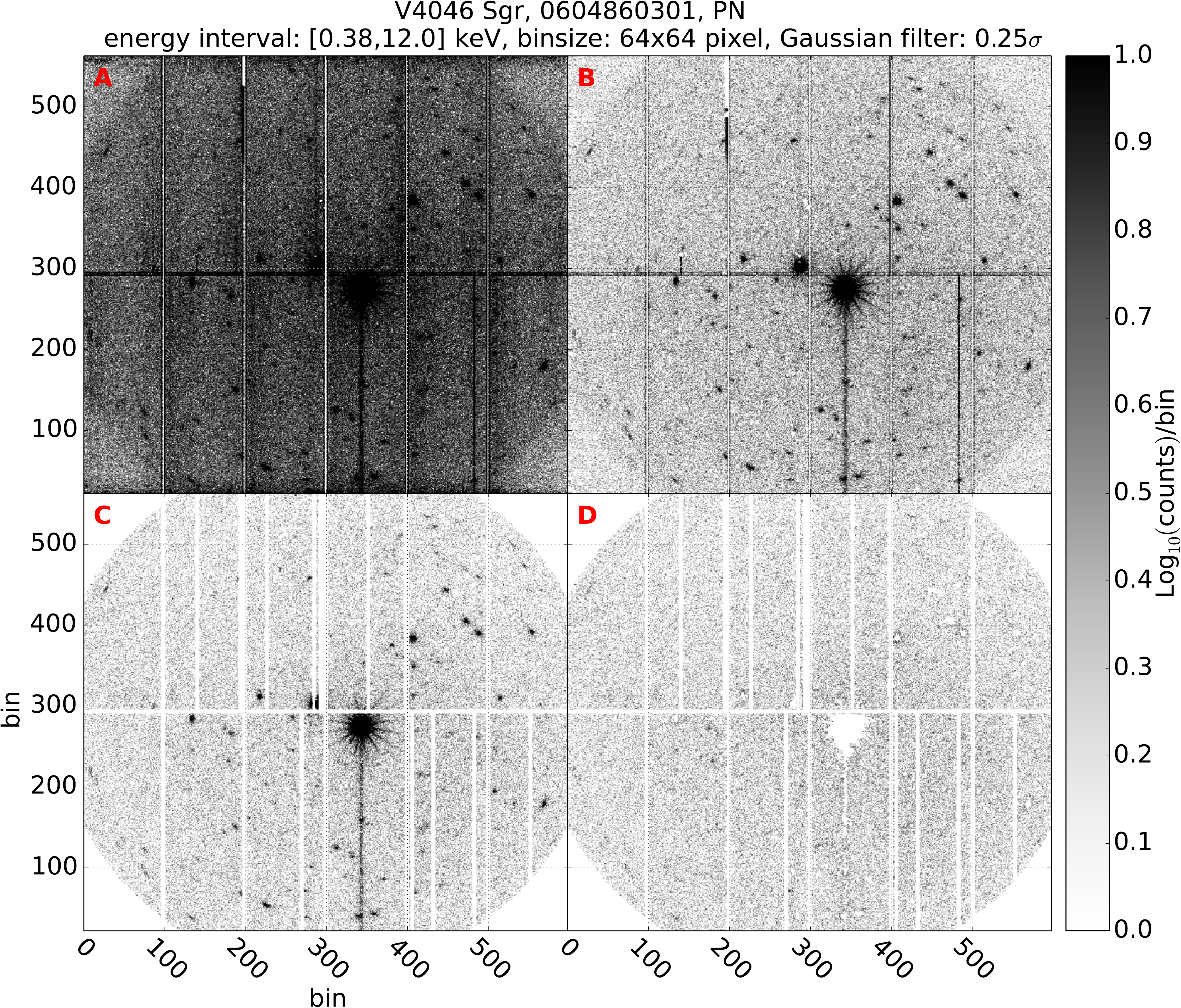}
\caption{All four panels show event files taken by the PN detector in an energy range of $0.38$ to $12.0$\,keV. The colour coding represents the number of X-ray photon counts per bin. The coordinats X and Y stand for the sky pixel coordinates. The upper left panel (A) shows the unfiltered raw event file and the upper right panel (B) is the time and particle background filtered result. The lower left panel (C) shows the application of the spatial filter and the lower right panel (D) represents the final source filtered diffuse X-ray background.}
\label{fig:TOP-TEN-0604860301-PN-4-bkg-380-12000-event-files26112015}
\end{figure}

The basic data reduction and analysis up to the generation of count spectra and their companion files have been achieved with the Scientific Analysis System (SAS 14.0.0)\footnote{"Users Guide to the XMM-Newton Science Analysis System", Issue 11.0, 2014 (ESA: XMM-Newton SOC).} and the related Current Calibration Files (CCF)\footnote{The Current Calibration Files are dated at February $3$rd, $2015$.} provided by the XMM-Newton Science Operations Center (XMM-SOC).
The SAS data reduction and analysis can be classified into four major steps: Reprocessing, time filtering, source detection and spectral analysis. The reprocessing tasks have been applied with the default adjustments recommended by the SAS-team. The script $espfilt$ from the ESAS\footnote{The XMM-ESAS software package is based on the software used for the background modelling described in \cite{Snowden:2004ap}.} software package was used for the time filtering of the data sets. It filters flares in the light curves by cutting the tails of the count rate histogram gained from the light curve itself. The energy range has been set from $300$\,eV to $12,000$\,eV. For illustration, the light curves of the field of view and the corners of the data set $0604860301$ are plotted in figure \ref{fig:TOP-TEN-0604860301-PN-4-bkg-380-12000-lightcurve-files26112015}. The figure shows the raw and filtered field of view and corner lightcurves, respectively. The light curves of the corner area of the PN-CCD show smaller count rate levels than the field of view lightcurves which is mainly an effect scaling with the investigated detector area. The difference of the raw and filtered light curves comprises flares originating in instrumental background effects.
The source detection algorithm $vtpdetect$\footnote{See ``The Detect Reference Manual`` (December 2006) for further information: http://cxc.harvard.edu/ciao/download/doc/detect\_manual/} \cite{Ebeling:1993zz} included in the CIAO analysis software \cite{2006SPIE.6270E..1VF} provided by the Chandra X-ray Center was applied in our analysis. The $vtpdetect$ algorithm determines complex regions following isocontours of the detected sources instead of circular regions as the standard SAS source detection does. This approach results in to a reduced contamination by the exclusion of sources. The energy range was set from $380$\,eV to $16,500$\,eV during the source detection. The parameter $coarse$ defines the lower threshold of events to be interpreted as a real source and was set to $2$ instead of $10$. The number of false source events is the product of the parameter $limit=10^{-6}$ times the number of background events. Note, the choice of parameters is driven by the need to keep the bias of the background due to unresolved sources low at the expense of misidentified sources.\\
The SAS source detection procedure was executed in addition to obtain an upper limit of the remaining source photons in the X-ray background, with some changes to the default parameters of the tasks. The energy range was set from $500$ to $4,000$\,eV for all tasks involved in the SAS source detection. The task $emask$ masks all pixels of the detector which have an exposure below the fraction of the maximum exposure denoted by the parameter $threshold\it{1}$. The latter was set to $0.1$ instead of the default value of $0.3$ to avoid excessive rejection of pixels on the detector and therefore photon counts in the event file. The minimum detection likelihood $likemin$ or $mlmin$ was adjusted down to $3.0$ from $10.0$ for the tasks $eboxdetect$, $emldetect$ and $esensmap$. The number of detection runs $nruns$ of the command $eboxdetect$ was increased from $3$ to $4$. This change was necessary to identify very faint or weak sources since the focus of this analysis lay on the diffuse X-ray background. Furthermore, the source selection radius $scut$ and the source cut-out radius $ecut$ of the script $emldetect$ were adjusted to $0.4$ and $0.95$, respectively. The latter two parameters represent the encircled energy as a fraction of the calibration point spread function. The column denoted by $N_{\text{source}}^{\text{X-ray}}$ in table \ref{tab:continuum} lists upper limits of the percentage of photons from the cut sources which possibly remained in the filtered background event file and are calculated as: $N_{\text{source}}^{\text{X-ray}}=\frac{100\,(1-ecut)\,N_{\text{source}}}{N_{\text{background}}}$. This upper limit also applies to the $vtpdetect$ results. The source detection algorithms can be compared in Appendix A.\\
The spectral analysis was applied with changes in the parameters of the tasks $rmfgen$ and $arfgen$: The energy interval was expanded from $0.05$\,keV to $20.48$\,keV and the number of energy bins was increased to a range of $30$ to $4096$ bins to match the spectral resolution of the physical energy channels of the PN-detector.\\
The figure \ref{fig:TOP-TEN-0604860301-PN-4-bkg-380-12000-event-files26112015} visualises the raw event file (upper left panel) and the outcomes of the three successive analysis steps: the time and particle background filtering (upper right panel), the spatial filtering (lower left panel) and the source filtering (lower right panel) of the data set $0604860301$.

\subsection{Instrumental Background}

\begin{figure}[ht]
\includegraphics[width=1.0\textwidth]{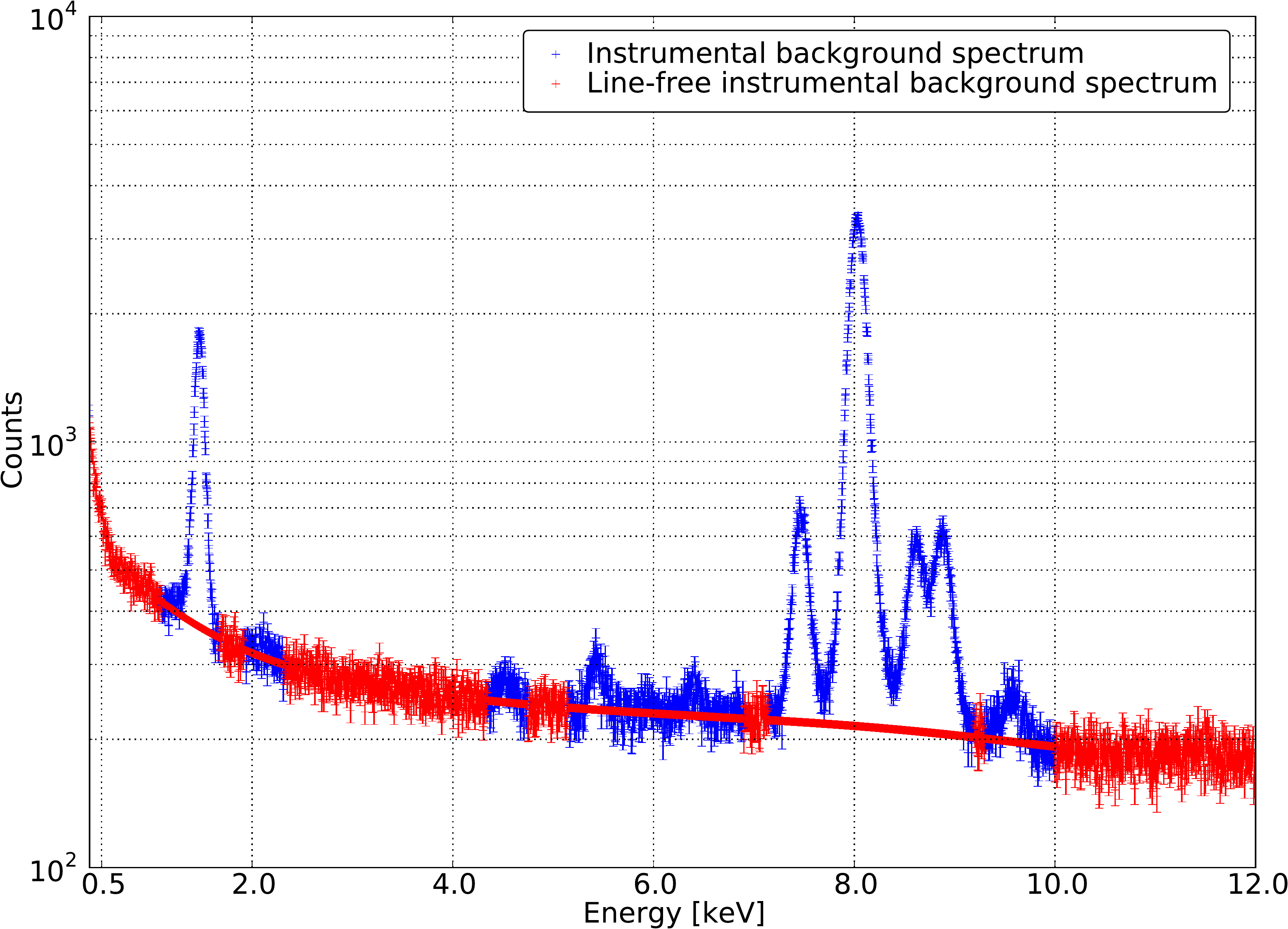}
\caption{The model of the instrumental background continuum shows the measured line-free segments (red crosses with error bars) and the fitted gaps (red line) which have been dominated by spectral lines in the primary spectrum. A polynomial of the $7$th order has been applied to bridge the gaps.}
\label{fig:instrumental}
\end{figure}

The data taken with the X-ray telescope XMM-Newton suffer from both non-negligible instrumental and particle background contamination and electronic noise \cite{Katayama:2002re}. The satellite is frequently exposed to energetic cosmic radiation, varying systematically in orbit as well as clouds of low energy protons entering the telescope. Figure \ref{fig:instrumental} presents the instrumental background spectrum and its model applied in this analysis. The instrumental background spectrum shows emission lines due to fluorescence effects triggered by cosmic-ray interactions with the support structure of the CCD-detectors and the detector material itself. The most dominant lines are identified as aluminum Al-K$_{\alpha}$ ($1.49$\,keV), nickel Ni-K$_{\alpha}$ ($7.48$\,keV), copper Cu-K$_{\alpha}$ ($8.05$\,keV and $8.95$\,keV), zirconium Zn-K$_{\alpha}$ ($8.64$\,keV and $9.57$\,keV) and molybdenum Mo-K$_{\alpha}$ ($17.4$\,keV). The filter-wheel-closed event files provided by the XMM-SOC were used to generate the spectra of the instrumental background for each detector. The event file of the instrumental background is a merger of all data taken while the filter-wheel of the telescope was closed\footnote{The corresponding event file in full-frame mode of the version of the year $2013$ was taken from the webpage of the XMM-SOC: http://xmm2.esac.esa.int/external/xmm\_sw\_cal/background/filter\_closed/pn/index.shtml.}. Despite the closed filter wheel, the light curve shows flares induced by cosmic ray incidents. Therefore, it has been time-filtered with the ESAS script $espfilt$ from the ESAS software package. This procedure led to the spectrum with a net exposure of $215\,\text{ks}$. The merged and time-filtered instrumental background continuum has been fitted\footnote{The method applied here is alike the fitting procedure used in the software package ESAS \cite{Snowden:2004ap}.} simultaneously in the energy intervals: $(0.65;1.10)\,\text{keV}$, $(1.675;1.925)\,\text{keV}$, $(2.325;4.350)\,\text{keV}$, $(4.75;5.15)\,\text{keV}$ and $(10.00;13.25)\,\text{keV}$ with a polynomial of the form
\begin{align}
 R(E)=\sum_{i=0}^{7}\,a_i\,\left(\frac{E}{\text{keV}}\right)^i.
\end{align}
The above energy intervals have been selected to avoid spectral lines in the fitted energy range. The resulting goodness-of-fit is $\chi^2=1324.82$ with $1338$ degrees of freedom (d.o.f.) and a probability value or p-value of $0.596$ and best-fit coefficients $\vec{a}=(739.57,\,-2.26, 4.79\cdot 10^{-3},\,-5.86\cdot 10^{-6},\,4.29\cdot 10^{-9},\,-1.85\cdot 10^{-12},\,4.28\cdot 10^{-16},\,-4.09\cdot 10^{-20})^{\text{T}}\,\text{s}^{-1}$.
The final model of the instrumental background continuum which is composed of line-free measured and fitted parts, is also plotted as red error bars in figure \ref{fig:instrumental}. Finally, the fitted instrumental background spectra have been normalised to the astrophysical background spectra of the energy range from $12.5$ to $14.0$\,keV under the assumption that the photon counts of astrophysical origins are vanishing in the non-instrumental or astrophysical spectra from which the modelled instrumental background spectrum was eventually subtracted. The normalisation factor
\begin{align}
N_{\,\text{INST}}=\frac{R_{\,\text{background}}}{R_{\,\text{instrumental}}}
\end{align}
with
\begin{align}
 R_{i \in\{\text{background,instrumental}\}}=\sum_{E^{\,i}_{\,\,\text{Photon}}} \begin{cases}
  1, &\text{if}\quad 12.5\,\text{keV}\leq E^{\,i}_{\,\,\text{Photon}}\leq 14.0\,\text{keV},\\
  0, &\text{elsewhere},
  \end{cases}
\end{align}
is listed in table \ref{tab:continuum} for each individual data set. The effective area is close to zero in this energy regime, so that the fraction of the instrumental background dominates above $10$\,keV in case of the last-mentioned astrophysical background spectra. The partially fitted and normalised instrumental background was then subtracted from the astrophysical background spectrum with the program XSPEC \cite{1996ASPC..101...17A}\footnote{The program XSPEC V11 \cite{1996ASPC..101...17A} has been used for the purpose of subtracting the individually scaled instrumental background spectra of the astrophysical spectra and of fitting the remaining spectra with multiplicative and/or additive combinations of predefined spectra.}.

\subsection{Fitting strategy}

The analysis focusses on possible additional emission lines in the diffuse X-ray background. This leads to a null hypothesis model $\bf{m}_{\bf{0}}$ of a continuum with superposed line emission. The continuum is modelled as a powerlaw and a bremsstrahlung component, both absorbed, denoted as ''wabs(powerlaw+bremss)``. The line emission describes the sum of $19$ astrophysical and instrumental emission lines of Gaussian shape:
\begin{align}
\bf{m}_{\bf{0}} = \text{wabs(powerlaw+bremss)} + \sum_{i=1}^{19}\,\text{Gaussian}.
\end{align}
The fitted Gaussian lines at energies of $0.58$\,keV (O VII), $0.67$ KeV (O VIII), $0.76$\,keV (Fe XVII), $0.84$\,keV (Fe XVII), $0.93$\,keV (Ne IX), $1.04$\,keV (Ne IX), $1.37$\,keV (Mg XI) and $2.34$\,keV (Si XIII) are interpreted as emission lines of  astrophysical origin while the instrumental Gaussian lines are located at energies of $1.49$\,keV (Al-K$_{\alpha}$), $1.87$\,keV (Si-K$_{\alpha}$), $4.54$\,keV (Ti-K$_{\alpha}$), $5.43$\,keV (Cr-K$_{\alpha}$), $6.42$\,keV (Fe-K$_{\alpha}$), $7.48$\,keV (Ni-K$_{\alpha}$), $8.04$\,keV (Cu-K$_{\alpha}$), $8.08$\,keV (Cu-K$_{\alpha}$), $8.63$\,keV (Cu-K$\beta$ and Zn-K$_{\alpha}$), $8.90$\,keV (Cu-K$_{\alpha}$) and $9.58$\,keV (Zn-K$_{\alpha}$). All $23$ data sets listed in table \ref{tab:datasets} have been fitted simultaneously in an energy range from $0.38$ to $12.0$\,keV by the application of a joint fitting procedure using the program XSPEC. The quality of the resulting best-fit reaches $\chi^2 = 52519.26$, $\text{d.o.f} = 52529$ and $\text{p-value} = 0.5111$. These values indicate a stable and good fit. No additional systematic uncertainties have been considered. The intrinsic widths of the astrophysical lines have been frozen to zero\footnote{A Gaussian line width of zero is interpreted as the width of one energy channel ($0.005$\,keV) by Xspec.} in contrast to the instrumental Gaussian emission lines. Some of the astrophysical lines have not been detected in all data sets.\\

\subsection{Statistical Methods}

The posterior predictive p-value analysis, in combination with the application of the $\mathcal{F}_{\text{Bevington}}$-statistics \cite{1969drea.book.....B} as test statistics, was applied as a hypothesis test of two models with differing degrees of freedom. The null hypothesis model $\bf{m}_{\bf{0}}$ consists of an absorbed bremsstrahlung continuum and a power law, which describe the background continuum as well as the $19$ additional Gaussian emission lines taking account of the instrumental and astrophysical emission lines. The alternative hypothesis $\bf{m}_{\bf{1}}$ is composed of the null hypothesis model $\bf{m}_{\bf{0}}$ plus an unidentified additive Gaussian emission line. A Gaussian emission line can be specified by three parameters: The normalisation, the width and the position of the mean of the Gausssian function on the energy axis. The width of the additional Gaussian line in the alternative hypothesis model has been set to zero in XSPEC so that the flux of the line was distributed over the energy resolution of one energy channel ($5$ eV) of the PN-detector. The forward-folding of the models with the related response matrices during the fitting process expands the line width to the energy-dependent spectral resolution. In order to compare the null-hypothesis model with the same model plus a Gaussian emission line as an additive component, two conditions have to be fulfilled \cite{Protassov:2002sz}:
\begin{enumerate}
\item The null hypothesis model $\bf{m}_{\bf{0}}$ and the alternative model $\bf{m}_{\bf{1}}$ must be nested so that the set of parameter values of one of the models is a subset of the set of parameter values of the other model.
\item The minimum values of the parameters of the additive model component are not allowed to lie on the boundary of the set of allowed parameter values.
\end{enumerate}
The first condition is true for the aforementioned hypothesis test, but the second condition is not obviously matched because the physically allowed minimum value of the flux of the additive emission line is zero. The $\mathcal{F}_{\text{Bevington}}$-statistics \cite{1969drea.book.....B} is defined as
\begin{align}
 \mathcal{F}_{\text{Bevington}}=\frac{\chi^2_0-\chi^2_1}{\frac{\chi^2_1}{\nu_1}}
\end{align}
and represents a measure to discriminate whether the null hypothesis model $\bf{m}_{\bf{0}}$ with its related goodness-of-fit $\chi^2_0$ or the alternative model $\bf{m}_{\bf{1}}$ with its related goodness-of-fit $\chi^2_1$ should be favoured. This is achieved by the difference of $\chi^2_0$ and $\chi^2_1$ divided by the reduced $\chi^2_1$, which is defined as $\chi^2_1$ divided by its number of degrees of freedom $\nu_1$. The alternative model $\bf{m}_{\bf{1}}$ will be preferred in the case of positive $\mathcal{F}_{\text{Bevington}}$-values ($\chi^2_0<\chi^2_1$) and vice versa. Nonetheless, the $\mathcal{F}_{\text{Bevington}}$-statistics is not a valid test statistics for the scenario mentioned above because of the aforementioned reasons \cite{Protassov:2002sz}. The $\mathcal{F}_{\text{Bevington}}$-statistics avoids these issues embedded in the posterior predictive p-value analysis as follows:
\begin{enumerate}
\item $M=500$ data sets are simulated from the null hypothesis model $\bf{m}_{\bf{0}}$ with the command 'fakeit' in XSPEC while using the original binning, effective areas, response matrices and background models.
\item The null hypothesis and the alternative models are simultaneously fitted to each of the $M$ simulated data sets and the corresponding Bevington-$\mathcal{F}$-values are calculated from the resulting $\chi^2_0$ and $\chi^2_1$ values \cite{1969drea.book.....B}, respectively. Each of the $23$ additive emission lines correspond to respectively each of the $23$ data sets (one per $i$th data set) in the alternative hypothesis model and are fitted for every point in a two-dimensional parameter grid which is determined by the energy and the flux normalisation $F_{\text{Photon}}$ of the additional Gaussian emission line. The flux normalisation $F_{\text{Photon}}$ of the $i$th additional line in the alternative model $\bf{m}_{\bf{1}}$ has been weighted by the factor
\begin{align}
w_i=\frac{S_{\,\text{NFW};i}\cdot t_{\,\text{exp};i}\cdot \Omega_{\,\text{fov};i}}{\sum_{j=1}^{N} \left(S_{\,\text{NFW};j}\cdot t_{\,\text{exp};j}\cdot \Omega_{\,\text{fov};j}\right)},\quad \sum_i\,w_i = 1, \quad i\in\left\{1,\ldots,N\right\} 
\end{align}
to take the NFW model-dependent dark matter column density, the field of view and the exposure of the individual observation into account. These weightings are listed in table \ref{tab:continuum}. The size of the grid ranges from $0.38$ to $12.0\,\text{keV}$ on a linear energy axis and from a total flux of $10^{-9}$ to $10^{3}\,\text{photons}\,\text{cm}^{-2}\,\text{s}^{-1}$ on a logarithmic normalisation axis. The resolution is $151\times 151$ grid points, the resulting energy resolution is approximately $76.95$ eV. 
\item The approximate q-value (defined as $q=1-p$, $p$ being the p-value) of a grid point $(E_{\gamma;m},F_{\,\text{Photon};n}), n,m\in\left\{1,\ldots,151\right\}$, is determined by adding up all $\mathcal{F}$-values of the simulated $\mathcal{F}$-distributions up to the measured $\mathcal{F}$-value as
\begin{align}
 &q(E_{\gamma;m},F_{\,\text{Photon};n})\\&=1-\frac{\sum^{M=500}_{k=1} \text{Ind}\left(\mathcal{F}_{\mathbf{Bevington}}^{\,\mathbf{Monte-Carlo};k}(E_{\gamma;m};F_{\,\text{Photon};n})>\mathcal{F}_{\mathbf{Bevington}}^{\,\mathbf{Measured}}(E_{\gamma;m};F_{\,\text{Photon};n})\right)}{M}\label{qvalue}.
\end{align}
The abbreviation ''$\text{Ind}$`` denotes the indicator function
\begin{align}
  &\text{Ind}\left(\mathcal{F}_{\mathbf{Bevington}}^{\,\mathbf{Monte-Carlo};k}(E_{\gamma;m};F_{\,\text{Photon};n})>\mathcal{F}_{\mathbf{Bevington}}^{\,\mathbf{Measured}}(E_{\gamma;m};F_{\,\text{Photon};n})\right)\\
  &=\begin{cases}
  1, &\text{if}\quad \mathcal{F}_{\mathbf{Bevington}}^{\,\mathbf{Monte-Carlo};k}(E_{\gamma;m};F_{\,\text{Photon};n})>\mathcal{F}_{\mathbf{Bevington}}^{\,\mathbf{Measured}}(E_{\gamma;m};F_{\,\text{Photon};n}), \\
  0, &\text{if}\quad \mathcal{F}_{\mathbf{Bevington}}^{\,\mathbf{Monte-Carlo};k}(E_{\gamma;m};F_{\,\text{Photon};n})\leq\mathcal{F}_{\mathbf{Bevington}}^{\,\mathbf{Measured}}(E_{\gamma;m};F_{\,\text{Photon};n}).
  \end{cases}
\end{align}
\end{enumerate}
The $\Delta\chi^2$-statistics was rejected as a hypothesis test but not as a method to obtain upper limits in the present work because of its inherent mathematical issues in the context to test for additive model components.
The value of $\Delta\chi^2$ is defined as the difference between a $\chi^2$ value and the local minimum in the $\chi^2$ space, both contained in the $l$th region of a number of $N_l,\,l \in \{1,...,N_l\}$ detected closed regions in which the nullhypothesis model $\bf{m}_{\bf{0}}$ was rejected by the posterior predictive p-value analysis, so that a $2\sigma$ confidence level has to fulfill the condition
\begin{align}
 \Delta\chi^2_l(E_{\gamma;m},F_{\,\text{Photon};n})_{2\sigma} = \chi^2_l(E_{\gamma;m},F_{\,\text{Photon};n}) - \min\left(\chi^2_l(E_{\gamma;m},F_{\,\text{Photon},n})\right) = 4.
\end{align}
Finally, the resulting set of $\Delta\chi^2_l(E_{\gamma;m},F_{\,\text{Photon};n})_{2\sigma}$ values has been evaluated to determine the $2\sigma$ confidence limits of the line energies $E_{\gamma}$ and the fluxes $F_{\,\text{Photon}}$ of the closed regions in which the null hypothesis model was rejected.
The combined application of the posterior predictive p-value analysis as a hypothesis test of additional emission lines and, in case of a rejection of the nullhypothesis model $\bf{m}_{\bf{0}}$, the successive determination of two-dimensional flux limits with help of the $\Delta\chi^2$-statistics, proved to be a powerful tool to uncover spectral emission lines.

\subsection{Combined model-dependent upper limits}

In some dark matter models particles with a mass $m_{\text{dm}}$ are proposed which are theoretically allowed to undergo two-body-decays to generate X-ray photons with an energy of $E_{\gamma}=\frac{m_{\text{dm}}}{2}$ in natural units with $m_{\text{dm}}$ being the mass of the dark matter particle \cite{Pal:1981rm,Dodelson:1993je,Abazajian:2012ys}. The decay measure of such a process contains the decay width $\Gamma_{\text{dm}}$ as the inverse of the decay time and is defined as
\begin{align}
 \epsilon_{\text{dm}}=\frac{1}{4\pi}\frac{E_{\gamma}\Gamma_{\text{dm}}}{m_{\text{dm}}}.
\end{align}
The expected intensity $I$ is the product of the decay measure and the model-dependent dark matter column density $S_{\text{dm}}$ which is an integral of a dark matter density distribution $\rho_{\text{dm}}$ over the distance $s$:
\begin{align}
 I_{\text{dm}}(s)=\epsilon_{\text{dm}}S_{\text{dm}}(s)=\frac{\Gamma_{\text{dm}}}{8\pi}\int_0^{\infty}\,\rho_{\text{dm}}(s)\,ds.
\end{align}
The upper limits of the flux have been used to constrain the parameter spaces of two theoretically proposed dark matter particles, the sterile neutrino and the axion. Both particles could decay into X-ray photons.
The decay rate of the Majorana sterile neutrino \cite{Pal:1981rm,Barger:1995ty} as a warm dark matter particle of mass $m_{\nu_s}$ is
\begin{align}
 \Gamma(\nu_R\rightarrow \gamma\nu_L)=\Gamma_{\nu_s}=\frac{9\alpha G_{\text{F}}^2}{1024 \pi^4} \sin^2(2\Theta) m_{\nu_s}^5\approx1.38\cdot 10^{-32}\,\text{s}^{-1}\left(\frac{\sin^2(2\Theta)}{10^{-10}}\right)\left(\frac{m_{\nu_s}}{\text{keV}}\right)^5,
\end{align}
with $\alpha$ being the fine structure constant and $G_{\text{F}}$ denotes the Fermi constant.
The Dirac sterile neutrino would have half the decay rate of the Majorana sterile neutrino.
The decay rate of an axion, which would be a cold dark matter particle \cite{Arias:2012az,Jaeckel:2014qea} with mass $m_{\phi}$ is
\begin{align}
  \Gamma(\phi\rightarrow\gamma\gamma)=\Gamma_{\phi}=\frac{64\pi}{g^2_{\phi\gamma\gamma}\,m^3_{\phi}}\approx 7.69\cdot 10^{-26}\,\text{s}^{-1}\,\left(\frac{g_{\phi\gamma\gamma}}{10^{-10}\,\text{GeV}^{-1}}\right)^{2}\left(\frac{m_{\phi}}{\text{eV}}\right)^{3}.
\end{align}
The intensity is composed of the flux divided by the field of view (FOV) while the energy flux is the product of the normalisation $F_{\,\text{Photon}}$ (photons per cm$^2$ per s) of the additional emission line and the energy of the line $E_{\gamma}$:
\begin{align}
 I_{\text{dm}}=E_{\gamma}\cdot F_{\,\text{Photon}}\cdot\Omega_{\text{fov}}^{-1}.
\end{align}
The upper limits for the mixing angle $\sin^2(2\Theta)$ of the sterile neutrinos is constrained via the mixing angle by
\begin{align}
 \sin^2(2\Theta)\leq\left(\frac{E_{\gamma}\cdot F_{\text{Photon}}}{6.9\cdot 10^4\,\text{keV}\,\text{cm}^{-2}\,\text{s}^{-1}}\right)\left(\frac{S_{\text{dm}}}{10^{27}\,\text{keV}\,\text{cm}^{-2}}\right)^{-1}\left(\frac{\Omega_{\text{fov}}}{4\pi\,\text{str}}\right)^{-1}\left(\frac{m_{\nu_s}}{\text{keV}}\right)^{-5}.
\end{align}
The coupling of the axion is constrained by
\begin{align}
  g^2_{\phi\gamma\gamma}\leq\left(\left(\frac{E_{\gamma}\cdot F_{\text{Photon}}}{3.8\cdot 10^{30}\,\text{keV}\,\text{cm}^{-2}\,\text{s}^{-1}}\right)\left(\frac{S_{\text{dm}}}{10^{27}\,\text{keV}\,\text{cm}^{-2}}\right)^{-1}\left(\frac{\Omega_{\text{fov}}}{4\pi\,\text{str}}\right)^{-1}\left(\frac{m_{\phi}}{\text{keV}}\right)^{-3}\right)\,\text{GeV}^{-2}.
\end{align}
The combined upper limits of the coupling $\sin^2(2\Theta)$ can be calculated from the combined upper limits of the flux normalisation.
The total coupling between active and sterile neutrinos is
\begin{align}
 \sin^2(2\Theta)_{\text{total}}=\sum_{i=1}^{N}\left(\frac{w_i\cdot E_{\gamma}\cdot F_{\,\text{Photon};i}}{6.9\cdot 10^{4}\,\text{keV}\,\text{cm}^{-2}\,\text{s}^{-1}}\right)\left(\sum_{j=1}^{N}\left(\frac{S_{\,\text{NFW};j}}{10^{27}\,\text{keV}\,\text{cm}^{-2}}\cdot\frac{\Omega_{\,\text{fov};j}}{4\pi}\right)\right)^{-1}\left(\frac{2 E_{\gamma}}{\text{keV}}\right)^{-5}\label{sin22theta}.
\end{align}
The very same procedure leads to the combined upper limit of the coupling constant of the axions:
\begin{align}
  &g^2_{\phi\gamma\gamma,\text{total}}=\\
  &\left(\sum_{i=1}^{N}\left(\frac{w_i\cdot E_{\gamma}\cdot F_{\,\text{Photon};i}}{3.8\cdot 10^{30}\,\text{keV}\,\text{cm}^{-2}\,\text{s}^{-1}}\right)\left(\sum_{j=1}^{N}\left(\frac{S_{\,\text{NFW};j}}{10^{27}\,\text{keV}\,\text{cm}^{-2}}\cdot\frac{\Omega_{\,\text{fov};j}}{4\pi}\right)\right)^{-1}\left(\frac{2 E_{\gamma}}{\text{keV}}\right)^{-3}\right)\,\text{GeV}^{-2}\label{gphi}.
\end{align}

\section{Results}

\subsection{Instrumental and astrophysical lines}

\begin{figure}
\includegraphics[width=1.0\textwidth]{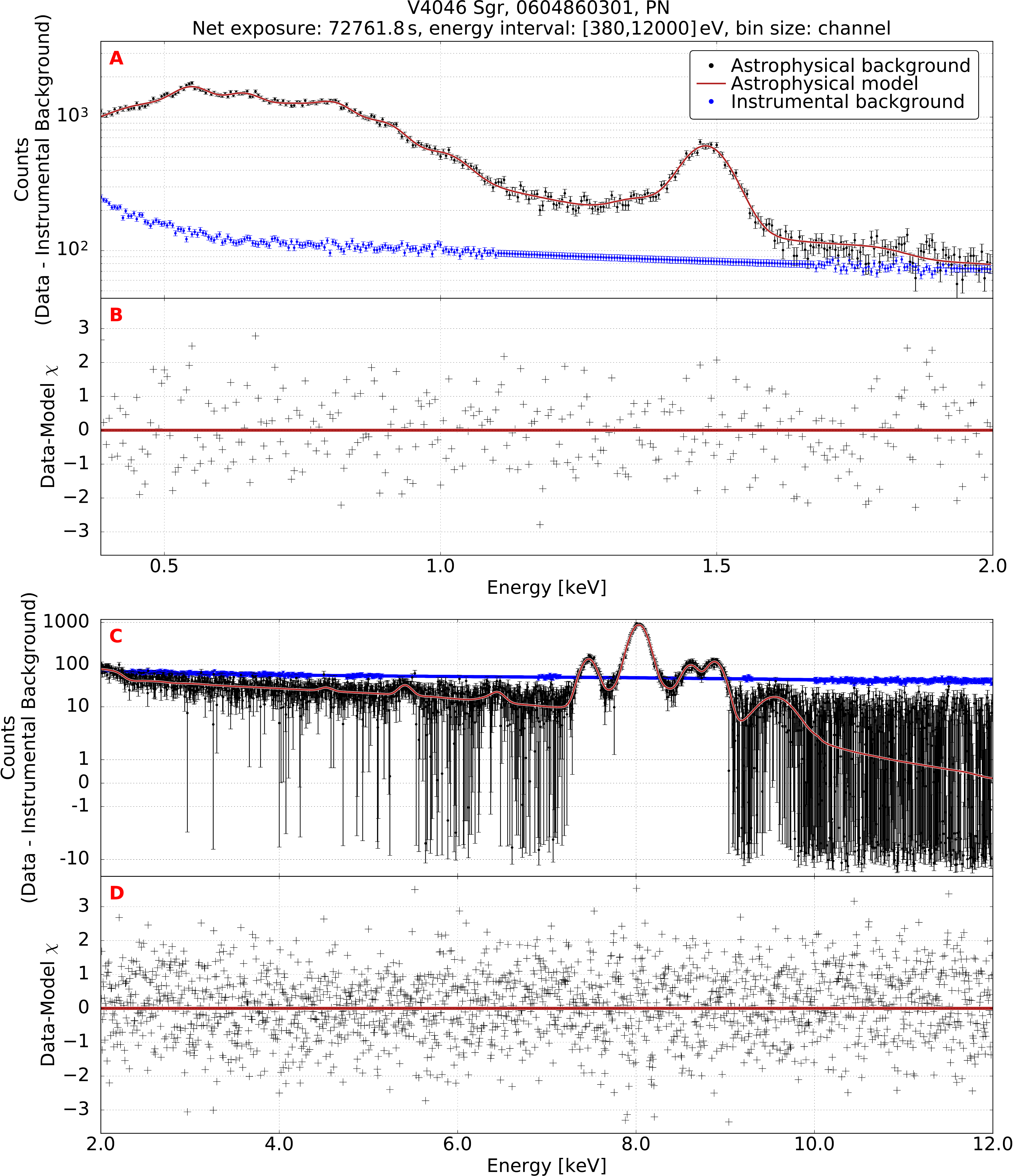}
\caption{\label{fig:model_single} The panels A and B show the measured spectrum of an individual data set ($0604860301$) and its fitted model in an energy range from $0.38$ to $2.0$\,keV (upper panel) and from $2.0$ to $12.0$\,keV (lower panel), respectively. The fitted model is composed of two continuum components consisting of a bremsstrahlung and a powerlaw component, both attenuated by absorption, plus $19$ instrumental and astrophysical Gaussian emission lines. The best-fit values are specified in tables \ref{tab:continuum} and \ref{tab:astrophysical}. The panels B and D show the residuals per energy bin in values of $\chi=\text{sgn}(\text{Data}-\text{Model})\,\sqrt{(\text{Data}-\text{Model})^2}$.}
\label{fig:TOP-TEN-0604860301-PN-0-bkg-380-12000-spectral-files26112015}
\end{figure}

\begin{figure}[ht]
\includegraphics[width=1.0\textwidth]{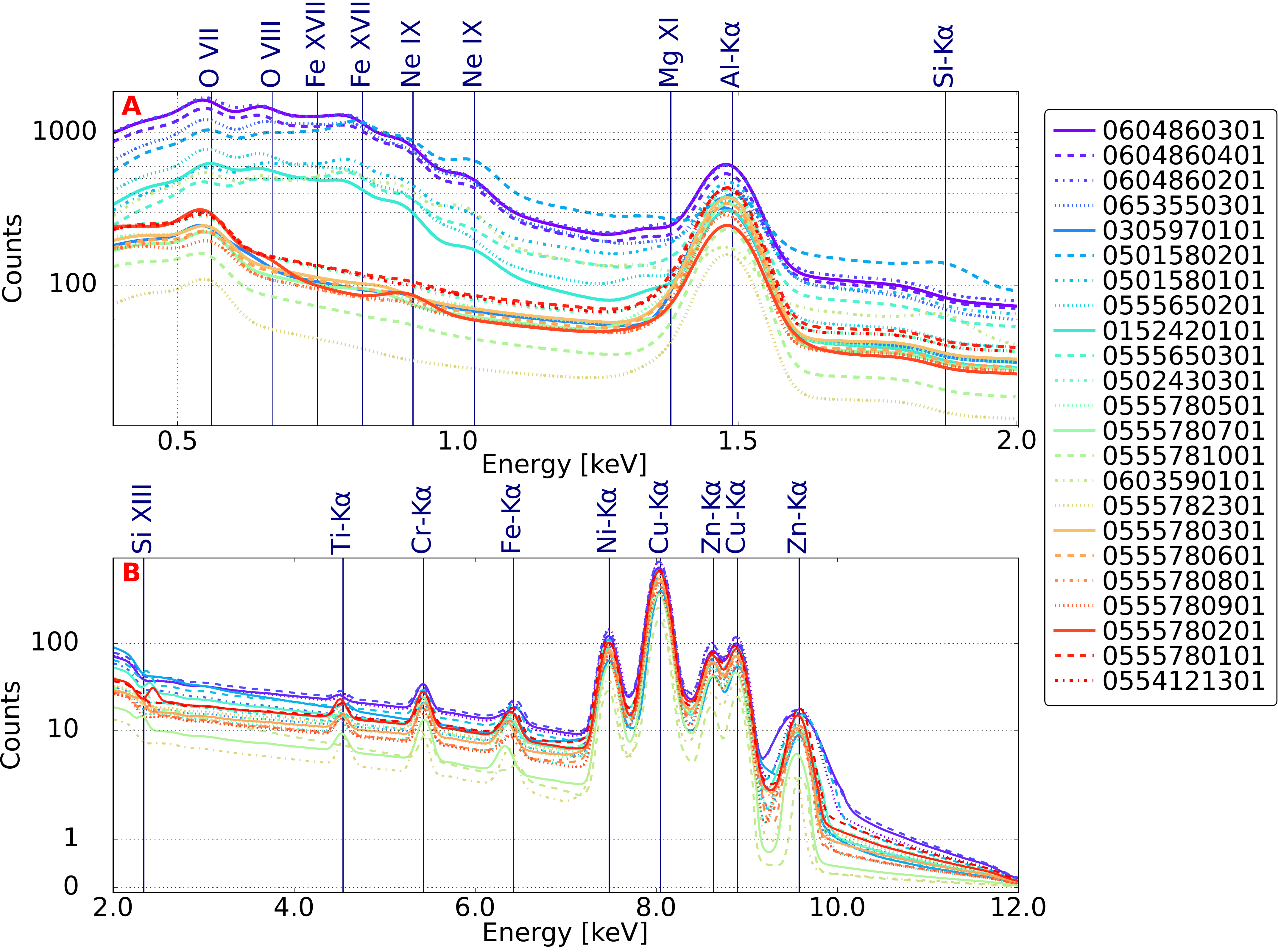}
\caption{\label{fig:model} The panels A and B shows the modelled spectra of the $23$ data sets in energy ranges of $0.38$\,keV to $2.0$\,keV (upper panel) and of $2.0$\,keV to $12.0$\,keV (lower panel), respectively. The fitted model consists of two continuum components consisting of a powerlaw and a bremsstrahlung component, both attenuated by absorption, plus $19$ instrumental and astrophysical Gaussian emission lines. The best-fit values are specified in tables \ref{tab:continuum} and \ref{tab:astrophysical}.}
\label{fig:DATA-SETS-380-12000-residuals-files}
\end{figure}

\begin{figure}[ht]
\includegraphics[width=0.99\textwidth]{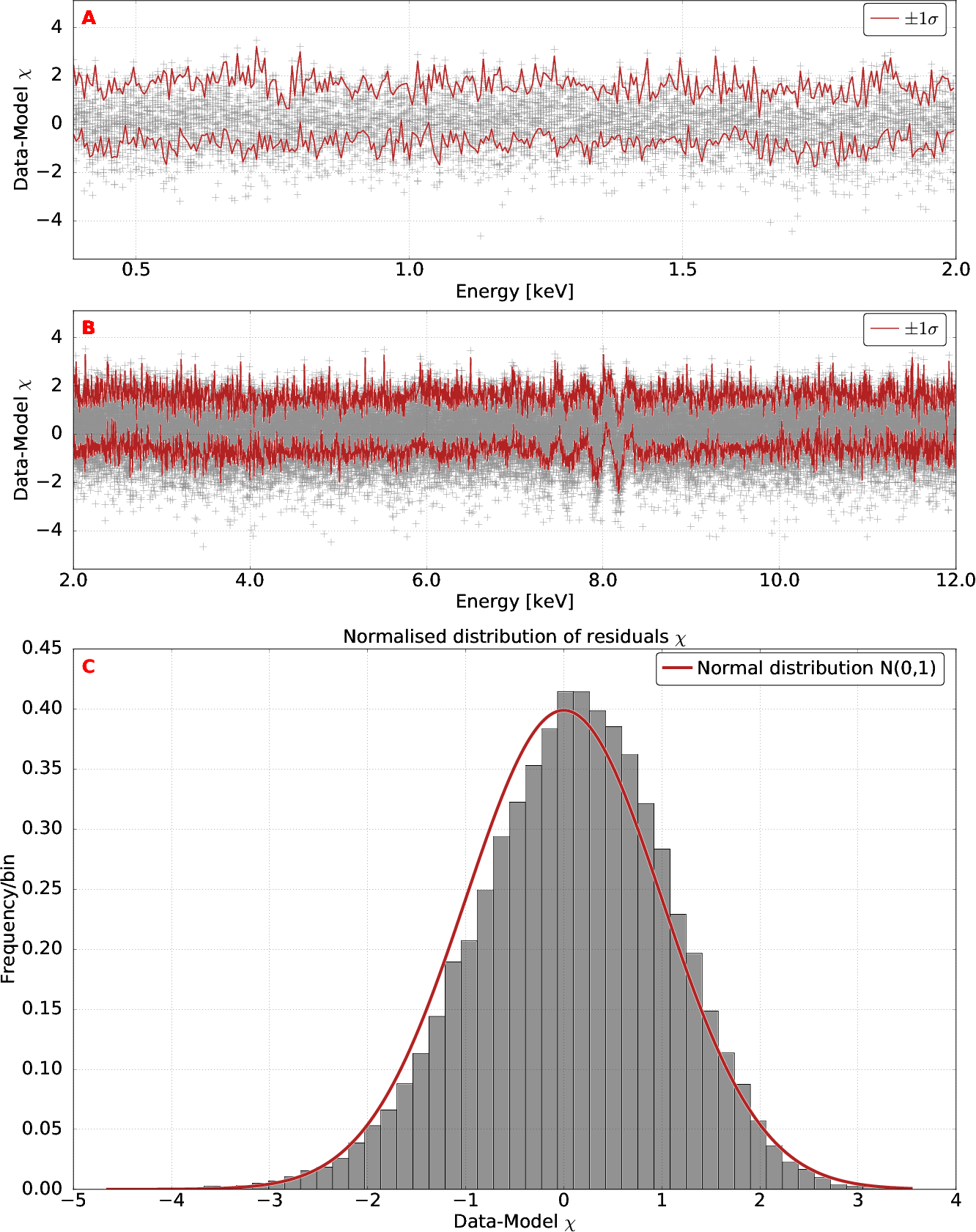}
\caption{The panels A and B show the residuals $\chi$ among all best-fit models and their underlying data sets in energy regimes of $0.38$\,keV to $2.0$\,keV (upper panel) and $2.0$\,keV to $12.0$\,keV (lower panel), respectively. The red curves indicate the $\pm 1\sigma$ levels of the residual distributions in each energy bin. The panel C presents the accumulated residual distribution of all energy bins. A normal distribution is also plotted for comparison.}
\label{fig:DATA-SETS-380-12000-residuals-files2}
\end{figure}

The instrumental background subtracted spectrum of the data set $0604860301$, the related best-fit model, the normalised instrumental background spectrum and the spectrum of the detected sources are plotted in the panels A and C of figure \ref{fig:TOP-TEN-0604860301-PN-0-bkg-380-12000-spectral-files26112015}. The x-axis is the energy scale in $\text{keV}$ and the y-axis indicates the photon counts. The panels B and D show the residuals among the model and the data in units of $\chi=\text{sgn}(\text{Data}-\text{Model})\,\sqrt{(\text{Data}-\text{Model})^2}$.\\
The resulting fitted values of the continuum component of the null hypothesis model are listed in table \ref{tab:continuum}. Table \ref{tab:astrophysical} contains the normalisations $F_{\text{Photon}}$ of the $19$ weighted Gaussian emission lines within the null hypothesis model. Normalisations have been ignored (denoted by "-" in table \ref{tab:continuum} and \ref{tab:astrophysical}) if the square root of the diagonal elements of the covariance matrix of the Levenberg-Marquardt fit were equal or higher than their central value. The absorption is given as the hydrogen column density $n_{\text{H}}$ and acts on the powerlaw and the bremsstrahlung component. The powerlaw is represented as the dimensionless photon index and its normalisation. The bremsstrahlung component is defined by the temperature in units of kT and its normalisation. The table \ref{tab:astrophysical} contains the mean energy and the normalisation values of the all Gaussian-shaped spectral emission lines with an astrophysical and an instrumental origin, respectively. The line energies are given as a mean value since their variations are in the order of $\mathcal{O}(eV)$. The photon fluxes of the instrumental lines are consistent with the overall activation of the telescope and its detectors by cosmic rays and soft proton clouds. The best-fit models of all $23$ data sets are plotted in figure \ref{fig:DATA-SETS-380-12000-residuals-files}. The panels A and B show the modelled counts per bin of the astrophysical background spectra in energy ranges from $0.38$\,keV to $2.0$\,keV and from $2.0$\,keV to $12.0$\,keV, respectively.\\
The residuals distributions per energy bin in figure \ref{fig:DATA-SETS-380-12000-residuals-files2}, panels A and B as well as the accumulated residual distribution of all energy bins, presented in figure \ref{fig:DATA-SETS-380-12000-residuals-files2}, panel C, have a recognisable shift of their means towards positive values. This bias of the residuals $\chi=\text{sgn}(\text{Data}-\text{Model})\,\sqrt{(\text{Data}-\text{Model})^2}$ can be explained by a non-vanishing contribution of photon counts from astrophysical sources in the normalisation range among the instrumental background model and the astrophysical spectra contrary to the assumption made in section 2.3. 

\subsection{Combined flux limits}

\begin{figure}[ht]
\includegraphics[width=1.0\textwidth]{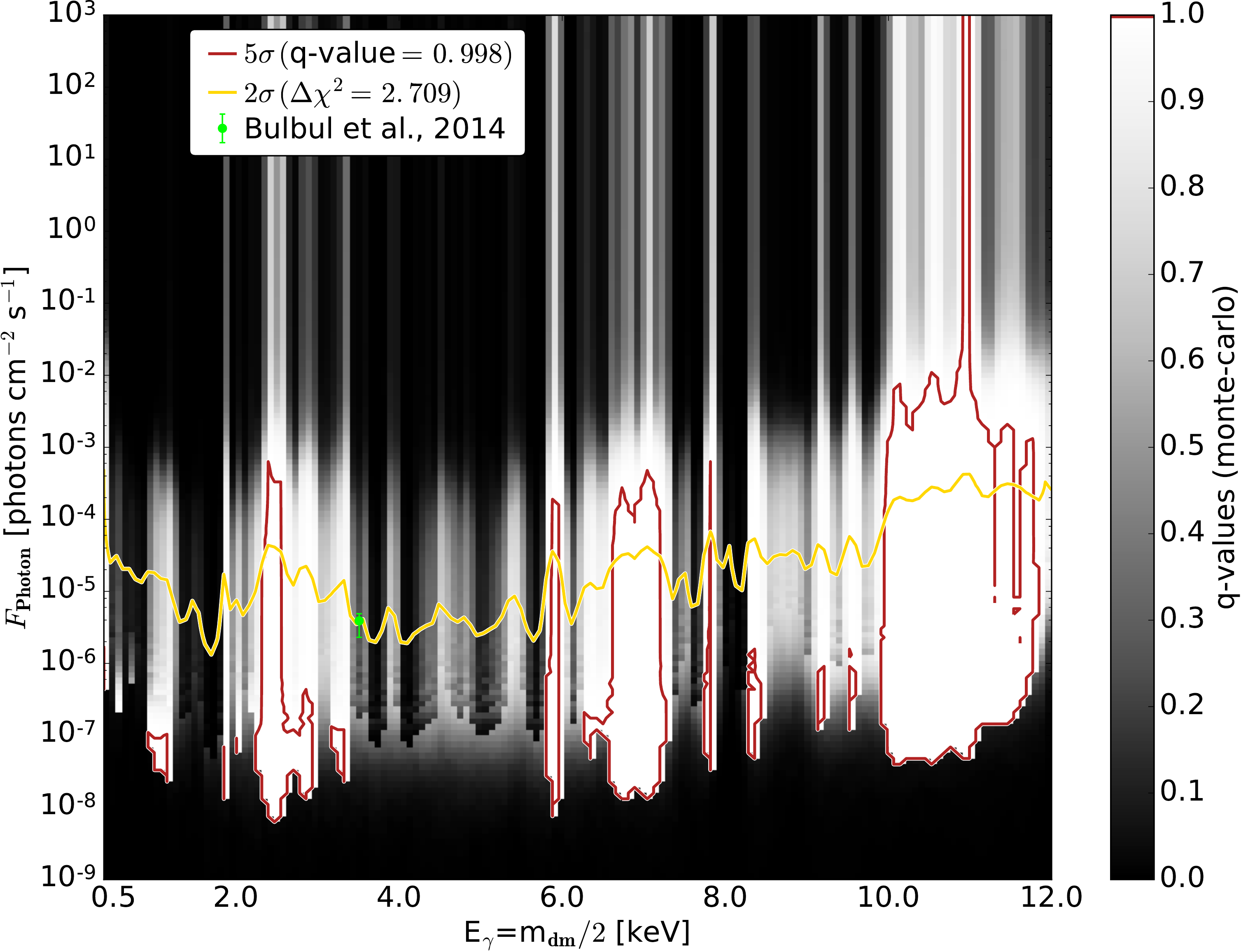}
\caption{The q-values and the combined $q=0.998$ confidence regions (red curve) on the total flux normalisation $F_{\text{Photon}}$ of an additional Gaussian emission line as the outcome of the posterior predictive p-value analysis. The energy regime ranges from $0.38\,\text{keV}$ to $12.0\,\text{keV}$. The colour coding and its related colour bar on the right next to the two-dimensional plot shows the level of the q-values. A $\Delta\chi^2$-value of $2.706$ is plotted for a comparison. The emission line detected by \cite{Bulbul:2014sua} at an energy of $3.51\pm 0.03$\,keV and a normalisation of $3.9^{\,+0.6}_{\,-1.0}\cdot 10^{-6}$\,cm$^{-2}$\,s$^{-1}$ is indicated by an orange error bar.}
\label{fig:NORM_UL}
\end{figure}

A q-value of $0.998$\footnote{$q=1-\frac{1}{M}=1-\frac{1}{500}=0.998$.} has been chosen to discriminate the parameter space of an additional emission line into regions in which the null hypothesis model $\bf{m}_{\bf{0}}$ is favoured, ($q\leq 0.998$), and in which the alternative model $\bf{m}_{\bf{1}}$ is preferred, ($q>0.998$). The figure \ref{fig:NORM_UL} shows the results of the posterior predictive p-value analysis. The panel contains the q-values calculated by equation \ref{qvalue} for every point $(E_{\gamma;m},F_{\,\text{Photon};n})$ in the two-dimensional grid spanned by the energy $E_{\gamma}$ and the normalisation $F_{\,\text{Photon}}$ of the additional emission line represented by the x-axis and the y-axis, respectively. The scaling of the colour coding is defined by the colour bar on the right side next to the panel. The contours of a q-value of $0.998$ are presented in the panel as red solid lines. A $\Delta\chi^2$-value of $2.706$ is plotted for a comparison (orange curve). The emission line detected by \cite{Bulbul:2014sua} at an energy of $3.51\pm 0.03$\,keV and a normalisation of $3.9^{\,+0.6}_{\,-1.0}\cdot 10^{-6}$\,cm$^{-2}$\,s$^{-1}$ is indicated by a green-coloured error bar.

\subsection{Constraints on $\sin^2(2\Theta)$ and $g_{\phi\gamma\gamma}$}

\begin{figure}[ht]
\includegraphics[width=1.0\textwidth]{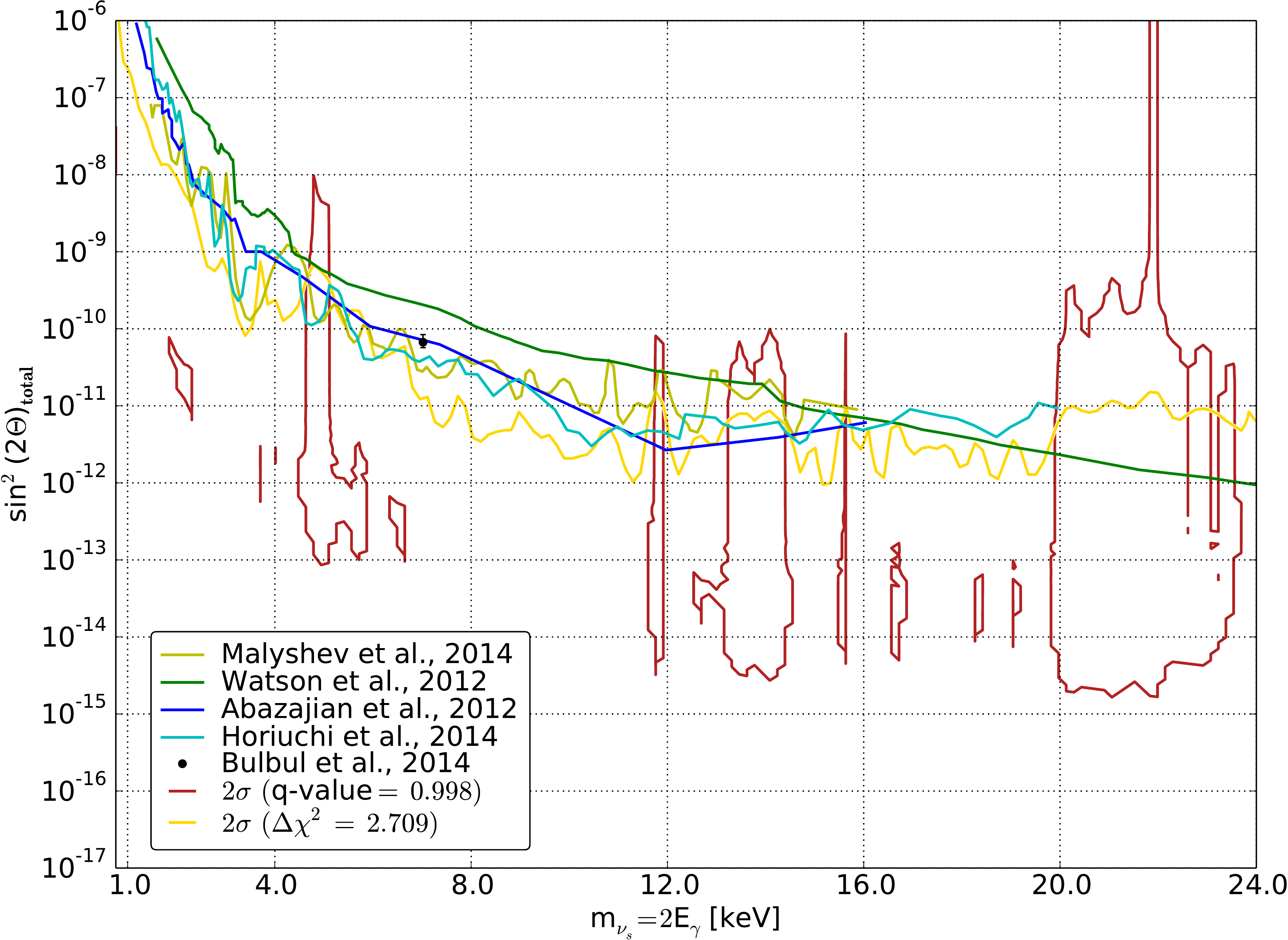}
\caption{The $q=0.998$ confidence regions of the active-sterile neutrino coupling $\sin^2(2\Theta)_{\text{total}}$ calculated from the $q=0.998$ confidence regions of the total flux normalisation $F_{\text{Photon}}$ in a sterile neutrino mass regime from $0.76\,\text{keV}$ to $24.0\,\text{keV}$. The results of this work are compared to the upper limit results of the publications \cite{Abazajian:2012ys,Malyshev:2014xqa,Horiuchi:2013noa,Watson:2011dw} and the detected emission line in \cite{Bulbul:2014sua}. A $\Delta\chi^2$-value of $2.706$ is plotted as a orange line for comparison. The emission line detected by \cite{Bulbul:2014sua} at an energy of $3.51\pm 0.03$\,keV and an active-sterile neutrino coupling of $6.7^{\,+1.7}_{\,-1.0}\cdot 10^{-11}$\,cm$^{-2}$\,s$^{-1}$ is indicated by a black error bar.}
\label{fig:SIN22THETA_UL}
\end{figure}

\begin{figure}[ht]
\includegraphics[width=1.0\textwidth]{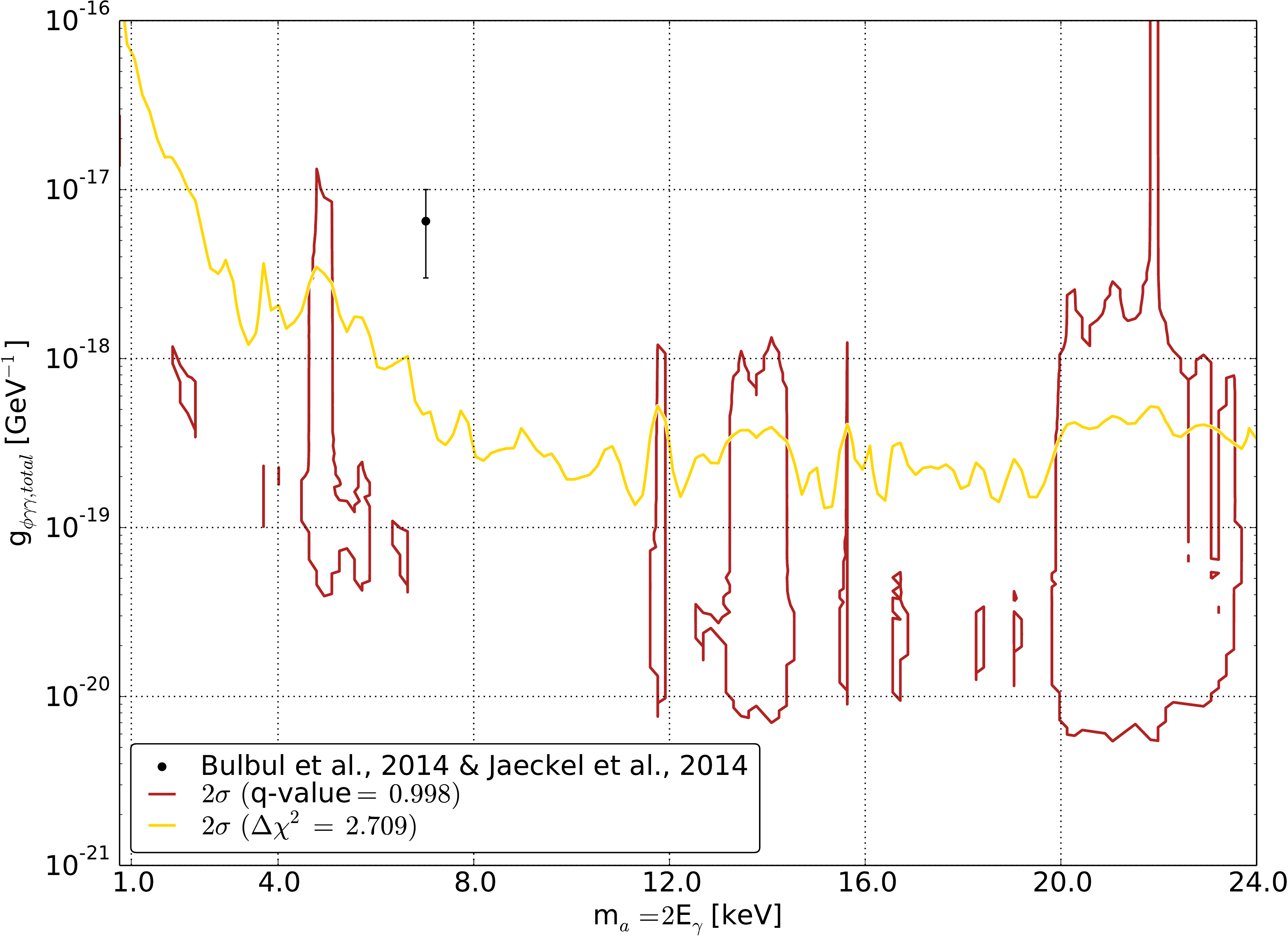}
\caption{The $q=0.998$ confidence regions of the active-sterile netrino coupling $g_{\phi\gamma\gamma,\text{total}}$ calculated from the upper limits of the total flux normalisation in an axion mass regime from $0.76\,\text{keV}$ to $24.0\,\text{keV}$. The results of this work are compared with the detected emission line in \cite{Bulbul:2014sua}. A $\Delta\chi^2$-value of $2.706$ is plotted as a orange line for comparison. The emission line detected by \cite{Bulbul:2014sua} at an energy of $3.51\pm 0.03$\,keV is indicated by a black error bar.}
\label{fig:AXION_UL}
\end{figure}

The equations \ref{sin22theta} and \ref{gphi} were used to calculate the $q=0.998$ confidence limits of the sterile-active neutrino and the axion-photon coupling from the upper limits of the total flux normalisation plotted in figure \ref{fig:NORM_UL} and are shown in figures \ref{fig:SIN22THETA_UL} and \ref{fig:AXION_UL}, respectively. The confidence limits of the couplings $\sin^2(2\Theta)_{\text{total}}$ and $g_{\phi\gamma\gamma,\text{total}}$ are compared with the outcomes of the publications \cite{Abazajian:2012ys,Malyshev:2014xqa,Horiuchi:2013noa,Watson:2011dw} and the spectral emission line detected by the authors of \cite{Bulbul:2014sua} in figures \ref{fig:SIN22THETA_UL} and \ref{fig:AXION_UL}, respectively. The emission line candidate is located at an photon energy of $3.51\pm 0.03$\,keV and results in a active-sterile neutrino coupling of $6.7^{\,+1.7}_{\,-1.0}\cdot 10^{-11}$\,cm$^{-2}$\,s$^{-1}$ and is denoted by a green errorbar in the plot \ref{fig:SIN22THETA_UL} and \ref{fig:AXION_UL}, respectively.

\subsection{Additional lines in the diffuse X-ray background}

The most dominant contour features have centroids in the areas closed by the $q=0.998$ contour regions with energies and normalisations presented in table \ref{tab:centroids}. The most-likely physical origins are also listed in table \ref{tab:centroids} which give classical physical explanations for all possible emission lines at the positions of the centroids. The $q=0.998$ regions at $0.38$\,keV and above $10.0$\,keV are not closed and could be consequences of badly constrained fits because of low effective areas at this energies.
\begin{table}[bt]
\centering\small\setlength\tabcolsep{3pt}
\renewcommand\arraystretch{1.4}
\scalebox{0.95}{
\begin{tabular}{cccccl}
\hline
\multicolumn{2}{c}{Posterior predictive p-value} & \multicolumn{3}{c}{$\Delta\chi^2$}  & Physical origin\\
\hline
$E_{\gamma}$ & $F_{\,\text{Photon}}$ & $E_{\gamma}$ & $F_{\,\text{Photon}}$ & $\Delta\chi^2_{\text{min}}$ &  \\
\hline
keV & $10^{-4}\,\text{cm}^{-2}\,\text{s}^{-1}$ & keV & $10^{-4}\,\text{cm}^{-2}\,\text{s}^{-1}$ & & \\
\hline\hline
$1.07$&$0.000590$& - & - & - & Ne X/Na-K$_{\alpha}$     \\
$1.85$&$0.000304$& - & - & - & Si-K, instrumental     \\
$2.01$&$0.000687$& - & - & - & Al XIII and/or Si XIV     \\
$2.54$&$0.918683$&$2.47_{-0.09}^{+0.06}$&$0.21_{-0.11}^{+0.09}$&$-16.6$& S XV at $2.45$\,keV     \\
 -    &    -     &$2.86_{-0.13}^{+0.08}$&$0.10_{-0.09}^{+0.09}$&$-4.5$& S XV at $2.88$\,keV \\
$3.26$&$0.000619$& - & - & - & Ar XVIII at $3.31$\,keV     \\
$5.88$&$0.166257$&$5.88_{-0.02}^{+0.04}$&$0.17_{-0.10}^{+0.09}$&$-11.8$& Mn, instrumental line      \\
$6.77$&$0.366257$&$6.81_{-0.20}^{+0.41}$&$0.14_{-0.12}^{+0.13}$&$-6.0$& Fe XXV at $6.7$\,keV     \\
 -    &    -     &$7.04_{-0.11}^{+0.10}$&$0.21_{-0.15}^{+0.12}$&$-8.0$& Influence of iron K-edge at $7.04$\,keV \\ 
$7.81$&$0.645020$&$7.82_{-0.04}^{+0.04}$&$0.30_{-0.26}^{+0.26}$&$-5.4$& Ni XXVII at $7.79$\,keV    \\
$8.35$&$0.007506$& - & - & - & Residuals of Zn-K$_{\alpha}$ and Cu-K$_{\alpha}$, instrumental   \\
$9.17$&$0.003528$& - & - & - & Au, instrumental     \\ 
$9.53$&$0.008391$& - & - & - & Residual of Zn-K$_{\alpha}$, instrumental    \\
$10.89$&    -     & $10.14_{-0.10}^{+0.11}$&$0.91_{-0.46}^{+0.58}$&$-14.1$& - \\ 
$10.89$&    -     & $10.53_{-0.08}^{+0.10}$&$1.32_{-0.61}^{+0.67}$&$-17.5$& - \\ 
$10.89$&   -     & $10.92_{-0.04}^{+0.09}$&$1.91_{-0.69}^{+0.99}$&$-24.0$& - \\ 
$11.30$&   -     & $11.38_{-1.36}^{+0.31}$&$1.32_{-1.19}^{+1.16}$&$-5.0$& - \\ 
$11.57$&   -     & $11.46_{-1.44}^{+0.23}$&$1.32_{-1.17}^{+1.32}$&$-5.0$& - \\ 
\hline
\end{tabular}
}
\caption{\label{tab:centroids} The table lists the centroidal energies $E_{\gamma}$ and normalisations $F_{\,\text{Photon}}$ of closed $q=0.998$ and $\Delta\chi^2=4$ contour regions as well as the possible physical origins. Only regions with $\Delta\chi^2$-values equal or lower than zero have been taken into account.}
\end{table}

\begin{figure}[ht]
\includegraphics[width=1.0\textwidth]{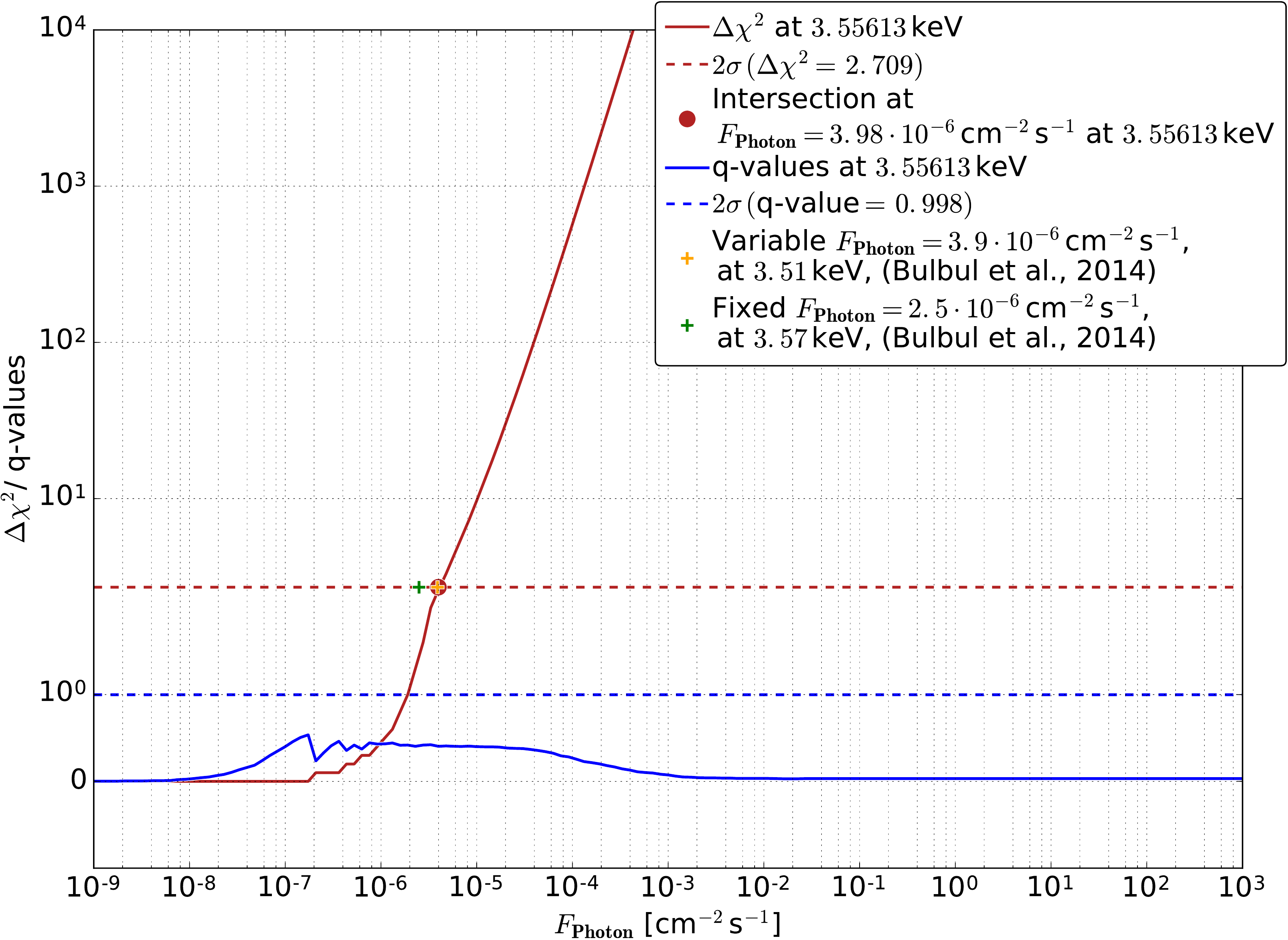}
\caption{This plot shows the trends of the $\Delta\chi^2$-values (red solid line) and the q-values (blue solid line) at an energy of $3.56$\,keV and a photon flux $F_{\text{Photon}}$ from $10^{-9}\,$cm$^{-2}$\,s$^{-1}$ to $10^{3}\,$cm$^{-2}$\,s$^{-1}$. The red and blue dashed line represent the $\Delta\chi^2=2.706$ and $q=0.998$ levels, respectively. The emission lines detected by \cite{Bulbul:2014sua} are denoted by a green and an orange crosses.}
\label{fig:DELTACHI2_COMBINED_26112015_150_0.38_12.0_150_1e-09_1000.0_gaussian_500_test_combined_NFW_3.55613}
\end{figure}

\begin{figure}[ht]
\includegraphics[width=1.0\textwidth]{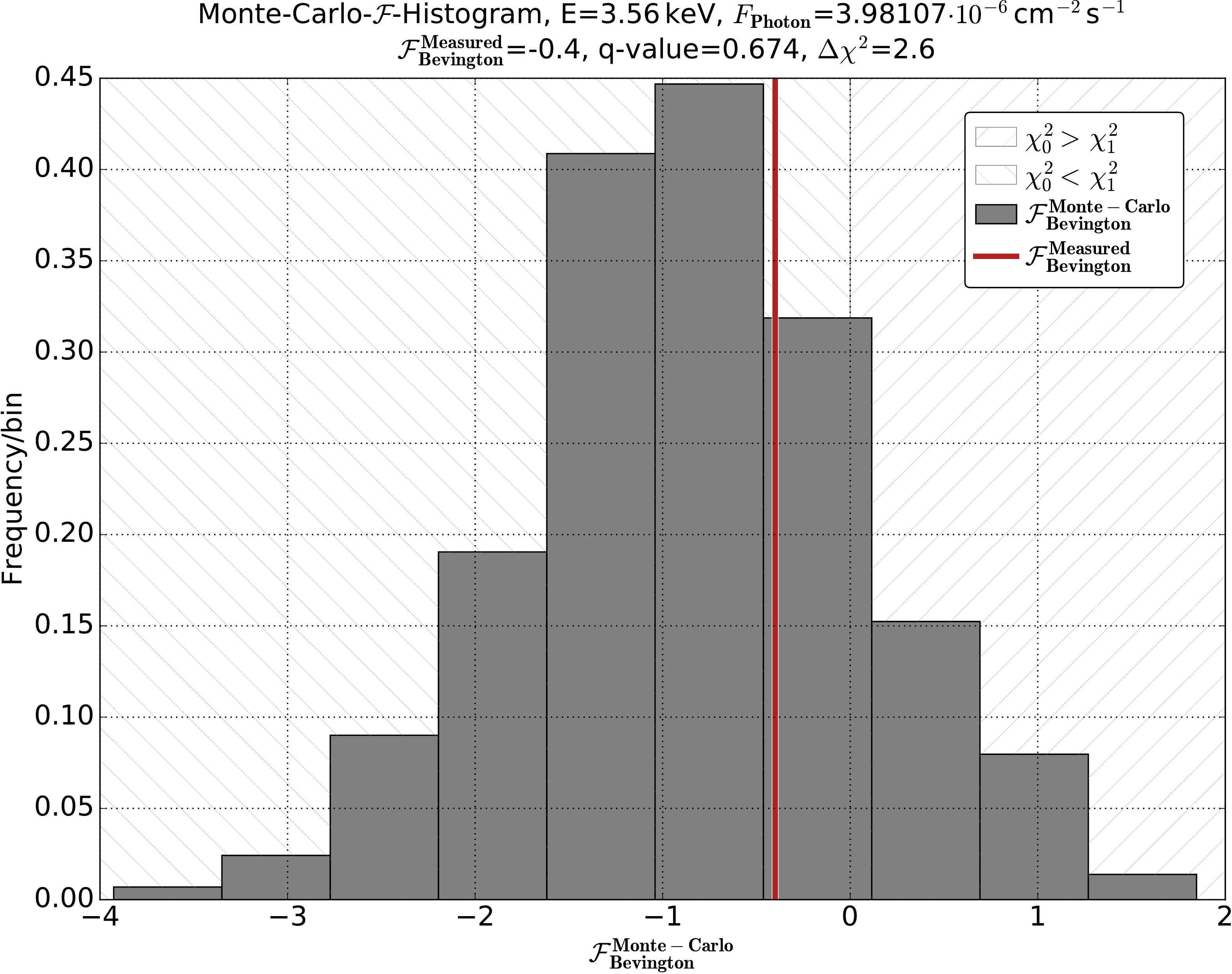}
\caption{The measured Bevington-$\mathcal{F}$-value $\mathcal{F}_{\mathbf{Bevington}}^{\,\mathbf{Measured}}(E_{\gamma}=3.56\,\text{keV};F_{\,\text{Photon}}=3.98\cdot 10^{-6}\,\text{cm}^{-2}\,\text{s}^{-1})$, denoted by the red line, and the distribution of the Monte-Carlo-$\mathcal{F}$-values $\mathcal{F}_{\mathbf{Bevington}}^{\,\mathbf{Monte-Carlo};k}(E_{\gamma}=3.56\,\text{keV};F_{\,\text{Photon}}=3.98\cdot 10^{-6}\,\text{cm}^{-2}\,\text{s}^{-1})$, denoted by the grey bars, are presented in this plot.}
\label{fig:SUM_F_HIST_COMBINED_26112015_150_0.38_12.0_150_1e-09_1000.0_gaussian_500_test_combined_NFW_3.98107e-06_3.55613_0.674}
\end{figure}

The authors of \cite{Bulbul:2014sua} claim an emission line at a photon energy of $3.51\pm 0.03$\,keV and a normalisation of $3.9^{\,+0.6}_{\,-1.0}\cdot 10^{-6}$\,cm$^{-2}$\,s$^{-1}$ in case of variable energies during their fit and a normalisation of $2.5^{\,+0.6}_{\,-0.7}\cdot 10^{-6}$\,cm$^{-2}$\,s$^{-1}$ for a fixed energy at $3.57$\,keV. The emission line is located in the $q\leq 0.998$ region of this work as shown in figure \ref{fig:DELTACHI2_COMBINED_26112015_150_0.38_12.0_150_1e-09_1000.0_gaussian_500_test_combined_NFW_3.55613}. The comparison of the measured Bevington-$\mathcal{F}$-value $\mathcal{F}_{\mathbf{Bevington}}^{\,\mathbf{Measured}}(E_{\gamma}=3.56\,\text{keV};F_{\,\text{Photon}}=3.98\cdot 10^{-6}\,\text{cm}^{-2}\,\text{s}^{-1})$ with the distribution of the Monte-Carlo-$\mathcal{F}$-values $\mathcal{F}_{\mathbf{Bevington}}^{\,\mathbf{Monte-Carlo};k}(E_{\gamma}=3.56\,\text{keV};F_{\,\text{Photon}}=3.98\cdot 10^{-6}\,\text{cm}^{-2}\,\text{s}^{-1})$ in figure \ref{fig:SUM_F_HIST_COMBINED_26112015_150_0.38_12.0_150_1e-09_1000.0_gaussian_500_test_combined_NFW_3.98107e-06_3.55613_0.674} do not reveal a significant preference towards the null hypothesis or the alternative model of our work. The q-value of $0.674$ illustrates that the measured $\mathcal{F}$-value lies close to the mean of the Monte-Carlo-$\mathcal{F}$-distribution. In summary, our results do not show any hints towards the existence of a spectral emission line at the energy claimed by \cite{Bulbul:2014sua}.
The centroids of $N_l = 12$ closed $q=0.998$ confidence regions with found in within this work (see table \ref{tab:centroids}) are located at energies which allow standard explanations of their physical origins. The most likely physical origins are also presented in table \ref{tab:centroids}. The centroid energies above $10.0$\,keV can be explained by the small effective area of the detector in that energy regime and the consequential low constraint on the fit. An $q=0.998$ region in which the null hypothesis model $\bf{m}_{\bf{0}}$ is rejected, overlaps with a corresponding $\Delta\chi^2$ area in which a minimum $\min\left(\chi^2_l(E_{\gamma;m},F_{\,\text{Photon};n})\right)$ exists as defined in section 2.5. The $2\sigma$ confidence levels of the $\Delta\chi^2$ statistics for $2$ degrees of freedom, expanded through the $E_{\gamma}$ and the $F_{\,\text{Photon}}$ parameter space, have been listed in table \ref{tab:centroids} as constraints on the minimal $\Delta\chi^2$ values $\min\left(\chi^2_l(E_{\gamma;m},F_{\,\text{Photon};n})\right)$. Unfortunately, it is not possible to resolve the photon fluxes of the centroids into the individual contributions of the $23$ data sets involved in the analysis because of the joint fitting and the consequential unified statistical handling of the data sets.

\section{Discussion}

A new type of analysis of X-ray background spectra has been carried out to search for unknown astrophysical or dark matter dependent emission lines. The application of a more rigorous and tailored statistical method in form of the posterior predictive p-value analysis was part of this work.
The NFW dark matter distribution as a model for the halo of our Milky Way was chosen to achieve a comparability with previous publications in this field. The choice of the dark matter distribution within our analysis is nearly unlimited.
The instrumental background continuum model was built from the merger of all filter-wheel-closed (FWC) data of the PN-detector available up to the year 2013 and scaled and subtracted from the astrophysical spectra. The subtraction of the instrumental background spectrum has its drawback in increasing the errors of the counts per energy bin of the resulting spectrum because of the necessary propagation of errors. It is important to note that the FWC data set used in this analysis is a merger of several single data sets from observation dates distributed over the operating duration of the satellite XMM-Newton. Therefore, the utilised FWC spectrum represents a time average of the energy distribution of the instrumental background.
Unfortunately, the corner spectra were not applicable due to low numbers of counts because of the small chip area involved and additional instrumental emission lines which are not present in the field of view spectra.
An alternative method would have been the combined fitting of the spectra by astrophysical and instrumental models in a single step. It is important to note that the instrumental model would not be allowed to be folded by the response matrix during the fitting process in such a scenario.
The results presented here were also achieved by the posterior predictive p-value analysis proposed by \cite{Protassov:2002sz} which is well fitted to find confidence regions in parameter spaces of additional model components by the exertion of a hypothesis test. This allows to probe the parameter space of an additive component like a Gaussian curve for unidentified emission lines in spectra. The confidence regions found in the energy-flux parameter space of an additional line have been transferred to two popular dark matter models, namely, the sterile neutrino and the axion, both under the assumption of the NFW dark matter halo distribution.

\section{Conclusion}
In summary our search for unknown emission lines in the diffuse X-Ray background exceed previous searches. We did not find new lines, in particular the famous $3.55$\,keV line has not been seen.

\section{Acknowledgements}
Our work was supported by the SFB 676 TP C2. AGP thanks Keith Arnaud and Craig Gordon and the SAS-team, especially Ignacio de la Calle, for many fruitful discussions related to XSPEC and the SAS software, respectively.

\clearpage

\section{Appendix A - SAS source detection and CIAO $vtpdetect$}

\begin{figure}[ht]
\includegraphics[width=1.0\textwidth]{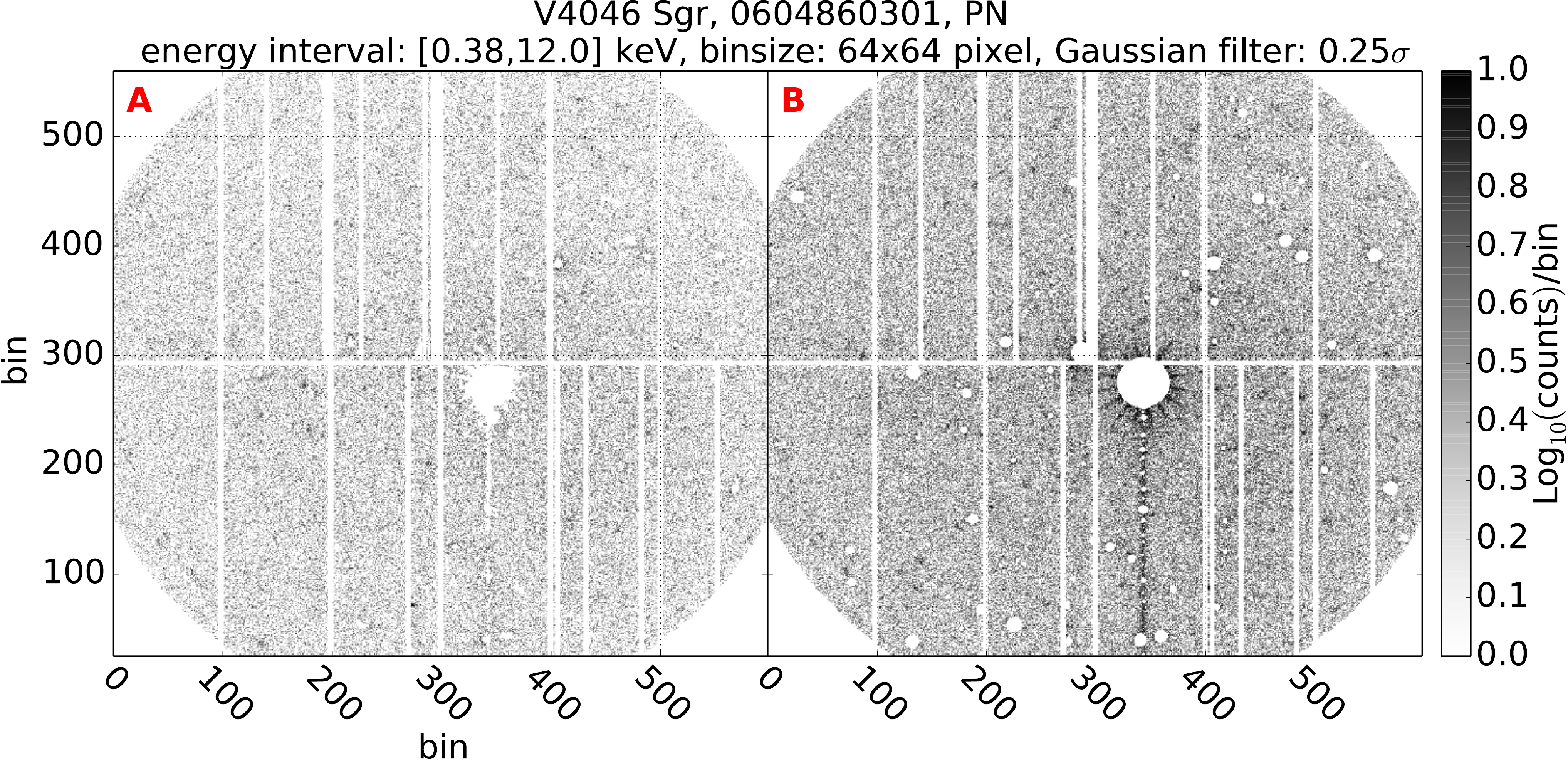}
\caption{The panels show two source-filtered event files of the data set $0604860301$ taken with the PN detector in an energy range of $0.38$ to $12.0$\,keV. The colour coding represents the number of X-ray photon counts per bin. The coordinates X and Y stand for the sky pixel coordinates. The right panel (B) represents the output of the source filtering routine by the SAS and the left panel (A) shows the output of the $vtpdetect$ source detection taken from the CIAO software package.}
\label{fig:TOP-TEN-0604860301-PN-4-bkg-380-12000-event-files-source-detection26112015}
\end{figure}

\clearpage

\section{Appendix B - Instrumental background model}

\clearpage

\section{Appendix C - Global fit parameters}

\begin{table}[ht]
\centering\small\setlength\tabcolsep{3.0pt}
\renewcommand\arraystretch{1.6}
\scalebox{0.8}{
\begin{tabular}{ccccccccccccccccccccccccccc}
\hline
Object & Obsid & $w_{\,\text{NFW}}$ & $N^{\text{X-ray}}_{\text{src}}$ & $N_{\text{INST}}$& Absorption		&	Bremsstr.		&	Bremsstr. 	&	Powerlaw	&	Powerlaw	\\
	& & & 			&&	$\text{n}_{\text{H}}$	&			 kT	&	 normalisation	&	$\Gamma$	&	normalisation	\\
\hline
& &  & $\%$ &  & 	$10^{21}\,\frac{\text{atoms}}{\text{cm}^{2}}$		&	$10^{-1}\,\text{keV}$		&	$10^{-4}\,\text{cm}^{-3}$		&		&	$\frac{10^{-4}\,\text{photons}}{\text{cm}^{2}\,\text{s}}\,\text{at}\,1\,\text{keV}$			\\
\hline	
\hline
V4046 Sgr&$0604860201$&$0.10$&$1.18$&$0.19$&$1.59\pm 0.17$& $2.78\pm 0.15$& $60.80\pm 14.11$&
 $1.01\pm 0.04$& $14.66\pm 0.64$\\ 
V4046 Sgr&$0604860301$&$0.12$&$1.23$&$0.23$&$1.59\pm 0.16$& $2.78\pm 0.14$& $61.66\pm 13.45$&
 $1.04\pm 0.03$& $14.66\pm 0.63$\\ 
V4046 Sgr&$0604860401$&$0.12$&$1.17$&$0.22$&$1.60\pm 0.16$& $2.79\pm 0.15$& $59.81\pm 13.24$&
 $1.09\pm 0.04$& $13.75\pm 0.64$\\ 
V4633 Sgr&$0653550301$&$0.09$&$0.07$&$0.15$&$2.11\pm 0.19$& $3.02\pm 0.20$& $66.22\pm 16.90$&
 $1.19\pm 0.05$& $14.90\pm 0.97$\\ 
PDS 456&$0501580201$&$0.08$&$2.10$&$0.18$&$2.26\pm 0.30$& $3.43\pm 0.40$& $24.65\pm 9.84$&
 $1.05\pm 0.05$& $13.04\pm 0.97$\\ 
PDS 456&$0501580101$&$0.06$&$3.34$&$0.14$&$2.30\pm 0.33$& $3.47\pm 0.47$& $28.03\pm 12.77$&
 $1.11\pm 0.06$& $15.99\pm 1.37$\\ 
VV Sco&$0555650301$&$0.04$&$0.69$&$0.12$&$2.26\pm 0.30$& $2.41\pm 0.18$& $104.94\pm 40.87$&
 $1.00\pm 0.06$& $11.55\pm 0.81$\\ 
VV Sco&$0555650201$&$0.05$&$0.68$&$0.14$&$2.02\pm 0.27$& $2.44\pm 0.18$& $85.49\pm 31.04$&
 $1.03\pm 0.05$& $13.96\pm 0.80$\\ 
CNOC2 Field 1&$0603590101$&$0.02$&$0.61$&$0.10$&$0.56\pm 0.28$& $2.15\pm 0.21$& $24.04\pm 11.25$&
 $0.92\pm 0.05$& $14.68\pm 0.87$\\ 
OGLE 1999 BUL 32&$0152420101$&$0.05$&$0.02$&$0.05$&$2.59\pm 0.32$& $2.69\pm 0.26$& $98.80\pm 43.47$&
 $1.82\pm 0.06$& $32.09\pm 2.38$\\ 
MACHO 96 BLG 5&$0305970101$&$0.09$&$0.07$&$0.10$&$2.58\pm 0.19$& $3.12\pm 0.22$& $86.07\pm 22.45$&
 $1.55\pm 0.05$& $33.02\pm 1.94$\\ 
LH VLA 2&$0554121301$&$0.01$&$0.11$&$0.07$&$1.66\pm 0.49$& $1.83\pm 0.21$& $54.99\pm 39.35$&
 $1.09\pm 0.08$& $14.40\pm 1.32$\\ 
RXJ2328.8+1453&$0502430301$&$0.02$&$0.06$&$0.13$&$1.01\pm 0.31$& $2.63\pm 0.31$& $13.13\pm 6.39$&
 $0.80\pm 0.05$& $10.77\pm 0.65$\\ 
CDFS&$0555780101$&$0.01$&$0.17$&$0.10$&$1.09\pm 0.37$& $2.41\pm 0.34$& $18.23\pm 10.93$&
 $1.00\pm 0.07$& $11.20\pm 0.93$\\ 
CDFS&$0555780201$&$0.01$&$0.15$&$0.12$&$0.94\pm 0.32$& $2.46\pm 0.30$& $16.00\pm 8.25$&
 $0.91\pm 0.05$& $12.71\pm 0.82$\\ 
CDFS&$0555780301$&$0.02$&$0.14$&$0.14$&$0.87\pm 0.29$& $2.71\pm 0.33$& $13.04\pm 5.96$&
 $0.77\pm 0.05$& $10.76\pm 0.67$\\ 
CDFS&$0555780501$&$0.02$&$0.12$&$0.18$&$0.92\pm 0.26$& $2.89\pm 0.33$& $11.83\pm 4.81$&
 $0.70\pm 0.05$& $10.00\pm 0.61$\\ 
CDFS&$0555780601$&$0.02$&$0.11$&$0.14$&$1.33\pm 0.29$& $2.47\pm 0.26$& $22.52\pm 10.00$&
 $0.81\pm 0.05$& $11.24\pm 0.74$\\ 
CDFS&$0555780701$&$0.02$&$0.26$&$0.17$&$1.07\pm 0.26$& $2.49\pm 0.24$& $18.19\pm 7.40$&
 $0.79\pm 0.05$& $11.24\pm 0.63$\\ 
CDFS&$0555780801$&$0.02$&$0.13$&$0.14$&$1.21\pm 0.31$& $2.53\pm 0.28$& $19.10\pm 9.00$&
 $0.80\pm 0.05$& $11.23\pm 0.74$\\ 
CDFS&$0555780901$&$0.01$&$0.12$&$0.13$&$1.17\pm 0.33$& $2.21\pm 0.24$& $23.95\pm 12.35$&
 $0.93\pm 0.05$& $12.73\pm 0.82$\\ 
CDFS&$0555781001$&$0.02$&$0.16$&$0.17$&$1.08\pm 0.27$& $2.55\pm 0.27$& $17.51\pm 7.51$&
 $0.84\pm 0.05$& $12.77\pm 0.71$\\ 
CDFS&$0555782301$&$0.02$&$0.12$&$0.15$&$1.10\pm 0.29$& $2.51\pm 0.26$& $17.91\pm 7.86$&
 $0.78\pm 0.05$& $11.53\pm 0.69$\\ 

\hline
\end{tabular}
}
\caption{\label{tab:continuum} The table above lists the best-fit values of the absorbed bremsstrahlung and the powerlaw components.}
\end{table}

\begin{table}[ht]
\centering\small\setlength\tabcolsep{3.0pt}
\renewcommand\arraystretch{1.6}
\scalebox{0.8}{
\begin{tabular}{ccccccccccccccccccccccccccc}
\hline
&&&&$E_{\text{line}}\,[\text{keV}]$	&	0.58 	&	0.67 	&	0.75 	&	0.83 	&	0.92 	 \\
\hline
&&&&				Phys. origin	&	O VII	&	O VIII	&	Fe XVII	&	Fe XVII	& Ne IX \\
\hline
\hline
Object & Obsid & $w_{\,\text{NFW}}$ & $N^{\text{X-ray}}_{\text{src}}$ & $N_{\text{INST}}$&	$F_{\text{Photon}}$	&	$F_{\text{Photon}}$	&	$F_{\text{Photon}}$	&	$F_{\text{Photon}}$	&	$F_{\text{Photon}}$			\\
\hline
& &  & $\%$ &  &	$\frac{10^{-4}}{\text{cm}^{2}\,\text{s}}$	&	$\frac{10^{-4}}{\text{cm}^{2}\,\text{s}}$	&	$\frac{10^{-4}}{\text{cm}^{2}\,\text{s}}$	&	$\frac{10^{-4}}{\text{cm}^{2}\,\text{s}}$	&	$\frac{10^{-4}}{\text{cm}^{2}\,\text{s}}$	\\
\hline
\hline
V4046 Sgr&$0604860201$&$0.10$&$1.18$&$0.19$&$10.26\pm 0.24$& $6.44\pm 0.26$& $4.08\pm 0.25$&
 $5.39\pm 0.24$& $3.65\pm 0.15$\\ 
V4046 Sgr&$0604860301$&$0.12$&$1.23$&$0.23$&$10.72\pm 0.23$& $6.63\pm 0.25$& $4.10\pm 0.24$&
 $5.64\pm 0.23$& $3.60\pm 0.14$\\ 
V4046 Sgr&$0604860401$&$0.12$&$1.17$&$0.22$&$10.15\pm 0.22$& $6.28\pm 0.25$& $4.00\pm 0.24$&
 $5.39\pm 0.23$& $3.55\pm 0.15$\\ 
V4633 Sgr&$0653550301$&$0.09$&$0.07$&$0.15$&$9.59\pm 0.23$& $6.04\pm 0.29$& $5.32\pm 0.28$&
 $6.57\pm 0.25$& $4.09\pm 0.17$\\ 
PDS 456&$0501580201$&$0.08$&$2.10$&$0.18$&$3.85\pm 0.15$& $2.73\pm 0.19$& $2.30\pm 0.19$&
 $3.16\pm 0.18$& $2.08\pm 0.12$\\ 
PDS 456&$0501580101$&$0.06$&$3.34$&$0.14$&$4.47\pm 0.21$& $3.10\pm 0.23$& $2.90\pm 0.23$&
 $3.88\pm 0.20$& $2.26\pm 0.15$\\ 
VV Sco&$0555650301$&$0.04$&$0.69$&$0.12$&$8.46\pm 0.31$& $5.16\pm 0.38$& $3.44\pm 0.37$&
 $4.15\pm 0.36$& $2.89\pm 0.19$\\ 
VV Sco&$0555650201$&$0.05$&$0.68$&$0.14$&$9.68\pm 0.28$& $6.47\pm 0.36$& $3.99\pm 0.34$&
 $4.38\pm 0.32$& $3.08\pm 0.19$\\ 
CNOC2 Field 1&$0603590101$&$0.02$&$0.61$&$0.10$&$6.96\pm 0.31$& $1.14\pm 0.21$& $-$&
 $-$& $0.52\pm 0.12$\\ 
OGLE 1999 BUL 32&$0152420101$&$0.05$&$0.02$&$0.05$&$7.82\pm 0.30$& $4.07\pm 0.35$& $2.53\pm 0.42$&
 $4.96\pm 0.45$& $4.01\pm 0.23$\\ 
MACHO 96 BLG 5&$0305970101$&$0.09$&$0.07$&$0.10$&$8.09\pm 0.23$& $3.77\pm 0.28$& $3.70\pm 0.28$&
 $6.59\pm 0.26$& $4.57\pm 0.18$\\ 
LH VLA 2&$0554121301$&$0.01$&$0.11$&$0.07$&$3.83\pm 0.30$& $-$& $-$&
 $-$& $-$\\ 
RXJ2328.8+1453&$0502430301$&$0.02$&$0.06$&$0.13$&$2.95\pm 0.24$& $0.77\pm 0.23$& $-$&
 $-$& $0.15\pm 0.08$\\ 
CDFS&$0555780101$&$0.01$&$0.17$&$0.10$&$2.90\pm 0.23$& $-$& $-$&
 $-$& $-$\\ 
CDFS&$0555780201$&$0.01$&$0.15$&$0.12$&$2.60\pm 0.20$& $-$& $-$&
 $-$& $0.15\pm 0.09$\\ 
CDFS&$0555780301$&$0.02$&$0.14$&$0.14$&$2.78\pm 0.19$& $-$& $-$&
 $-$& $-$\\ 
CDFS&$0555780501$&$0.02$&$0.12$&$0.18$&$2.96\pm 0.17$& $-$& $-$&
 $-$& $-$\\ 
CDFS&$0555780601$&$0.02$&$0.11$&$0.14$&$2.91\pm 0.19$& $-$& $-$&
 $-$& $0.17\pm 0.09$\\ 
CDFS&$0555780701$&$0.02$&$0.26$&$0.17$&$2.92\pm 0.18$& $-$& $-$&
 $-$& $0.13\pm 0.08$\\ 
CDFS&$0555780801$&$0.02$&$0.13$&$0.14$&$3.20\pm 0.20$& $-$& $-$&
 $-$& $0.24\pm 0.10$\\ 
CDFS&$0555780901$&$0.01$&$0.12$&$0.13$&$3.49\pm 0.21$& $-$& $-$&
 $-$& $0.25\pm 0.09$\\ 
CDFS&$0555781001$&$0.02$&$0.16$&$0.17$&$3.41\pm 0.19$& $-$& $-$&
 $-$& $0.12\pm 0.08$\\ 
CDFS&$0555782301$&$0.02$&$0.12$&$0.15$&$2.99\pm 0.19$& $-$& $-$&
 $-$& $0.17\pm 0.09$\\ 

\hline
\multicolumn{10}{c}{To be continued on the next page.} \\
\hline
\end{tabular}
}
\end{table}

\begin{table}[ht]
\centering\small\setlength\tabcolsep{3.0pt}
\renewcommand\arraystretch{1.6}
\scalebox{0.8}{
\begin{tabular}{ccccccccccccccccccccccccccc}
\hline
&&&&$E_{\text{line}}\,[\text{keV}]$	&	1.03 &	1.38 	&	1.49	& 1.87 & 2.28\\
\hline
&&&&				Phys. origin	&	Ne IX&	Mg XI	&	Al-K$_{\alpha}$&Si-K&Si XIII\\
\hline
\hline
Object & Obsid & $w_{\,\text{NFW}}$ & $N^{\text{X-ray}}_{\text{src}}$ & $N_{\text{INST}}$&	$F_{\text{Photon}}$ &	$F_{\text{Photon}}$	&	$F_{\text{Photon}}$	&	$F_{\text{Photon}}$	& $F_{\text{Photon}}$ 	\\
\hline
& &  & $\%$ &  &	$\frac{10^{-4}}{\text{cm}^{2}\,\text{s}}$&	$\frac{10^{-4}}{\text{cm}^{2}\,\text{s}}$	&	$\frac{10^{-4}}{\text{cm}^{2}\,\text{s}}$	&	$\frac{10^{-4}}{\text{cm}^{2}\,\text{s}}$	&	$\frac{10^{-4}}{\text{cm}^{2}\,\text{s}}$	\\
\hline
\hline
V4046 Sgr&$0604860201$&$0.10$&$1.18$&$0.19$&$1.39\pm 0.09$& $0.40\pm 0.06$& $3.90\pm 0.08$& $-$&
 $-$\\ 
V4046 Sgr&$0604860301$&$0.12$&$1.23$&$0.23$&$1.44\pm 0.09$& $0.48\pm 0.05$& $3.89\pm 0.08$& $-$&
 $-$\\ 
V4046 Sgr&$0604860401$&$0.12$&$1.17$&$0.22$&$1.45\pm 0.09$& $0.44\pm 0.06$& $3.98\pm 0.08$& $-$&
 $-$\\ 
V4633 Sgr&$0653550301$&$0.09$&$0.07$&$0.15$&$1.91\pm 0.11$& $0.35\pm 0.06$& $3.28\pm 0.07$& $-$&
 $-$\\ 
PDS 456&$0501580201$&$0.08$&$2.10$&$0.18$&$0.78\pm 0.08$& $0.30\pm 0.06$& $3.37\pm 0.09$& $-$&
 $0.32\pm 0.05$\\ 
PDS 456&$0501580101$&$0.06$&$3.34$&$0.14$&$0.78\pm 0.10$& $0.28\pm 0.10$& $3.78\pm 0.09$& $-$&
 $0.25\pm 0.07$\\ 
VV Sco&$0555650301$&$0.04$&$0.69$&$0.12$&$0.77\pm 0.11$& $0.48\pm 0.10$& $4.61\pm 0.10$& $-$&
 $-$\\ 
VV Sco&$0555650201$&$0.05$&$0.68$&$0.14$&$0.97\pm 0.12$& $0.44\pm 0.07$& $4.52\pm 0.09$& $-$&
 $-$\\ 
CNOC2 Field 1&$0603590101$&$0.02$&$0.61$&$0.10$&$-$& $1.61\pm 0.35$& $4.89\pm 0.38$& $-$&
 $-$\\ 
OGLE 1999 BUL 32&$0152420101$&$0.05$&$0.02$&$0.05$&$2.51\pm 0.15$& $0.41\pm 0.08$& $1.72\pm 0.08$& $0.43\pm 0.07$&
 $-$\\ 
MACHO 96 BLG 5&$0305970101$&$0.09$&$0.07$&$0.10$&$2.80\pm 0.12$& $0.46\pm 0.07$& $2.29\pm 0.07$& $0.48\pm 0.06$&
 $-$\\ 
LH VLA 2&$0554121301$&$0.01$&$0.11$&$0.07$&$-$& $-$& $6.03\pm 1.01$& $-$&
 $-$\\ 
RXJ2328.8+1453&$0502430301$&$0.02$&$0.06$&$0.13$&$-$& $-$& $5.66\pm 0.10$& $-$&
 $-$\\ 
CDFS&$0555780101$&$0.01$&$0.17$&$0.10$&$-$& $0.31\pm 0.13$& $5.78\pm 0.16$& $-$&
 $0.26\pm 0.11$\\ 
CDFS&$0555780201$&$0.01$&$0.15$&$0.12$&$-$& $0.30\pm 0.12$& $5.78\pm 0.15$& $-$&
 $0.37\pm 0.10$\\ 
CDFS&$0555780301$&$0.02$&$0.14$&$0.14$&$-$& $0.44\pm 0.09$& $5.76\pm 0.11$& $-$&
 $0.27\pm 0.09$\\ 
CDFS&$0555780501$&$0.02$&$0.12$&$0.18$&$-$& $0.31\pm 0.07$& $6.01\pm 0.10$& $-$&
 $0.14\pm 0.08$\\ 
CDFS&$0555780601$&$0.02$&$0.11$&$0.14$&$-$& $0.41\pm 0.09$& $6.20\pm 0.12$& $-$&
 $0.25\pm 0.09$\\ 
CDFS&$0555780701$&$0.02$&$0.26$&$0.17$&$-$& $0.34\pm 0.07$& $6.15\pm 0.10$& $-$&
 $0.29\pm 0.09$\\ 
CDFS&$0555780801$&$0.02$&$0.13$&$0.14$&$-$& $0.24\pm 0.07$& $6.52\pm 0.11$& $-$&
 $0.47\pm 0.10$\\ 
CDFS&$0555780901$&$0.01$&$0.12$&$0.13$&$-$& $0.52\pm 0.14$& $6.03\pm 0.17$& $-$&
 $0.22\pm 0.10$\\ 
CDFS&$0555781001$&$0.02$&$0.16$&$0.17$&$-$& $0.51\pm 0.11$& $6.42\pm 0.13$& $-$&
 $0.34\pm 0.09$\\ 
CDFS&$0555782301$&$0.02$&$0.12$&$0.15$&$-$& $0.40\pm 0.13$& $6.26\pm 0.15$& $-$&
 $0.12\pm 0.09$\\ 

\hline
\multicolumn{10}{c}{To be continued on the next page.} \\
\hline
\end{tabular}
}
\end{table}

\begin{table}[ht]
\centering\small\setlength\tabcolsep{3.0pt}
\renewcommand\arraystretch{1.6}
\scalebox{0.8}{
\begin{tabular}{ccccccccccccccc}
\hline
&&&&$E_{\text{line}}\,[\text{keV}]$	 & 4.52 &	5.42	&	6.40  &	7.47 \\
\hline
&&&&				Phys. origin	   &  Ti-K$_{\alpha}$   & Cr-K$_{\alpha}$	 &	Fe-K$_{\alpha}$ 	&	Ni-K$_{\alpha}$  \\
\hline
\hline
Object & Obsid & $w_{\,\text{NFW}}$ & $N^{\text{X-ray}}_{\text{src}}$ & $N_{\text{INST}}$&	$F_{\text{Photon}}$	&	$F_{\text{Photon}}$	& $F_{\text{Photon}}$	&	$F_{\text{Photon}}$	\\
\hline
& &  & $\%$ &  &	$\frac{10^{-4}}{\text{cm}^{2}\,\text{s}}$	&	$\frac{10^{-4}}{\text{cm}^{2}\,\text{s}}$&	$\frac{10^{-4}}{\text{cm}^{2}\,\text{s}}$&	$\frac{10^{-4}}{\text{cm}^{2}\,\text{s}}$ \\
\hline
\hline
V4046 Sgr&$0604860201$&$0.10$&$1.18$&$0.19$&$0.08\pm 0.05$& $0.45\pm 0.05$& $0.22\pm 0.06$&
 $6.07\pm 0.19$\\ 
V4046 Sgr&$0604860301$&$0.12$&$1.23$&$0.23$&$0.10\pm 0.04$& $0.32\pm 0.05$& $0.29\pm 0.06$&
 $5.99\pm 0.17$\\ 
V4046 Sgr&$0604860401$&$0.12$&$1.17$&$0.22$&$0.11\pm 0.04$& $0.41\pm 0.05$& $0.28\pm 0.06$&
 $6.47\pm 0.18$\\ 
V4633 Sgr&$0653550301$&$0.09$&$0.07$&$0.15$&$0.20\pm 0.05$& $0.46\pm 0.05$& $0.31\pm 0.06$&
 $5.22\pm 0.17$\\ 
PDS 456&$0501580201$&$0.08$&$2.10$&$0.18$&$-$& $0.37\pm 0.05$& $0.33\pm 0.06$&
 $5.39\pm 0.16$\\ 
PDS 456&$0501580101$&$0.06$&$3.34$&$0.14$&$-$& $0.46\pm 0.06$& $0.25\pm 0.07$&
 $5.98\pm 0.27$\\ 
VV Sco&$0555650301$&$0.04$&$0.69$&$0.12$&$0.17\pm 0.06$& $0.52\pm 0.07$& $0.25\pm 0.09$&
 $6.98\pm 0.22$\\ 
VV Sco&$0555650201$&$0.05$&$0.68$&$0.14$&$0.14\pm 0.06$& $0.56\pm 0.07$& $0.28\pm 0.08$&
 $7.06\pm 0.23$\\ 
CNOC2 Field 1&$0603590101$&$0.02$&$0.61$&$0.10$&$0.33\pm 0.09$& $0.67\pm 0.10$& $0.37\pm 0.12$&
 $9.63\pm 0.37$\\ 
OGLE 1999 BUL 32&$0152420101$&$0.05$&$0.02$&$0.05$&$-$& $0.23\pm 0.05$& $0.08\pm 0.06$&
 $2.66\pm 0.15$\\ 
MACHO 96 BLG 5&$0305970101$&$0.09$&$0.07$&$0.10$&$0.08\pm 0.04$& $0.27\pm 0.05$& $0.30\pm 0.06$&
 $3.52\pm 0.13$\\ 
LH VLA 2&$0554121301$&$0.01$&$0.11$&$0.07$&$0.43\pm 0.12$& $0.86\pm 0.13$& $0.60\pm 0.15$&
 $9.36\pm 0.41$\\ 
RXJ2328.8+1453&$0502430301$&$0.02$&$0.06$&$0.13$&$0.21\pm 0.07$& $0.48\pm 0.08$& $0.30\pm 0.09$&
 $8.48\pm 0.24$\\ 
CDFS&$0555780101$&$0.01$&$0.17$&$0.10$&$0.24\pm 0.09$& $0.70\pm 0.10$& $0.27\pm 0.12$&
 $8.81\pm 0.32$\\ 
CDFS&$0555780201$&$0.01$&$0.15$&$0.12$&$0.31\pm 0.09$& $0.68\pm 0.10$& $0.25\pm 0.11$&
 $8.35\pm 0.31$\\ 
CDFS&$0555780301$&$0.02$&$0.14$&$0.14$&$0.21\pm 0.08$& $0.55\pm 0.09$& $0.33\pm 0.10$&
 $8.22\pm 0.29$\\ 
CDFS&$0555780501$&$0.02$&$0.12$&$0.18$&$0.24\pm 0.07$& $0.59\pm 0.08$& $0.53\pm 0.10$&
 $9.05\pm 0.27$\\ 
CDFS&$0555780601$&$0.02$&$0.11$&$0.14$&$0.23\pm 0.08$& $0.68\pm 0.09$& $0.39\pm 0.11$&
 $8.80\pm 0.30$\\ 
CDFS&$0555780701$&$0.02$&$0.26$&$0.17$&$0.26\pm 0.07$& $0.60\pm 0.08$& $0.40\pm 0.10$&
 $9.17\pm 0.30$\\ 
CDFS&$0555780801$&$0.02$&$0.13$&$0.14$&$0.33\pm 0.08$& $0.82\pm 0.10$& $0.55\pm 0.11$&
 $9.64\pm 0.33$\\ 
CDFS&$0555780901$&$0.01$&$0.12$&$0.13$&$0.35\pm 0.09$& $0.86\pm 0.10$& $0.59\pm 0.11$&
 $9.24\pm 0.33$\\ 
CDFS&$0555781001$&$0.02$&$0.16$&$0.17$&$0.38\pm 0.08$& $0.80\pm 0.09$& $0.49\pm 0.10$&
 $9.41\pm 0.27$\\ 
CDFS&$0555782301$&$0.02$&$0.12$&$0.15$&$0.26\pm 0.08$& $0.80\pm 0.09$& $0.51\pm 0.11$&
 $9.26\pm 0.31$\\ 

\hline
\multicolumn{8}{c}{To be continued on the next page.} \\
\hline
\end{tabular}
}
\end{table}

\begin{table}[ht]
\centering\small\setlength\tabcolsep{3.0pt}
\renewcommand\arraystretch{1.6}
\scalebox{0.8}{
\begin{tabular}{ccccccccccccccccccccccccccc}
\hline
&&&&$E_{\text{line}}\,[\text{keV}]$	&		8.03	&	8.19 &	8.57	&	8.88	&	9.57 \\
\hline
&&&&				Phys. origin	& Cu-K$_{\alpha}$	&	Cu-K$_{\alpha}$ &	Zn-K$_{\alpha}$	&	Cu-K$_{\alpha}$	&	Zn-K$_{\alpha}$		\\
\hline
\hline
Object & Obsid & $w_{\,\text{NFW}}$ & $N^{\text{X-ray}}_{\text{src}}$ & $N_{\text{INST}}$&	$F_{\text{Photon}}$	&	$F_{\text{Photon}}$	&	$F_{\text{Photon}}$	&	$F_{\text{Photon}}$	&	$F_{\text{Photon}}$		\\
\hline
& &  & $\%$ &  &	$\frac{10^{-4}}{\text{cm}^{2}\,\text{s}}$	&	$\frac{10^{-4}}{\text{cm}^{2}\,\text{s}}$	&	$\frac{10^{-4}}{\text{cm}^{2}\,\text{s}}$	&	$\frac{10^{-4}}{\text{cm}^{2}\,\text{s}}$	&	$\frac{10^{-4}}{\text{cm}^{2}\,\text{s}}$	\\
\hline
\hline
V4046 Sgr&$0604860201$&$0.10$&$1.18$&$0.19$&$51.25\pm 0.53$& $7.98\pm 0.51$& $7.36\pm 0.56$&
 $10.38\pm 0.32$& $4.22\pm 0.37$\\ 
V4046 Sgr&$0604860301$&$0.12$&$1.23$&$0.23$&$52.76\pm 0.50$& $7.91\pm 0.53$& $7.24\pm 0.33$&
 $10.82\pm 0.37$& $3.69\pm 0.33$\\ 
V4046 Sgr&$0604860401$&$0.12$&$1.17$&$0.22$&$53.19\pm 0.47$& $7.43\pm 0.69$& $7.63\pm 0.36$&
 $11.05\pm 0.37$& $3.10\pm 0.29$\\ 
V4633 Sgr&$0653550301$&$0.09$&$0.07$&$0.15$&$45.64\pm 0.41$& $6.76\pm 1.05$& $6.10\pm 0.52$&
 $9.21\pm 0.49$& $1.87\pm 0.27$\\ 
PDS 456&$0501580201$&$0.08$&$2.10$&$0.18$&$46.28\pm 0.48$& $6.50\pm 0.59$& $6.68\pm 0.54$&
 $8.93\pm 0.31$& $3.36\pm 0.31$\\ 
PDS 456&$0501580101$&$0.06$&$3.34$&$0.14$&$52.92\pm 0.54$& $7.10\pm 0.52$& $7.38\pm 1.05$&
 $10.13\pm 0.56$& $2.54\pm 0.35$\\ 
VV Sco&$0555650301$&$0.04$&$0.69$&$0.12$&$58.81\pm 0.90$& $7.87\pm 0.63$& $9.00\pm 0.60$&
 $11.28\pm 0.40$& $2.89\pm 0.43$\\ 
VV Sco&$0555650201$&$0.05$&$0.68$&$0.14$&$61.09\pm 0.84$& $8.16\pm 0.65$& $9.54\pm 0.92$&
 $12.05\pm 0.51$& $3.06\pm 0.38$\\ 
CNOC2 Field 1&$0603590101$&$0.02$&$0.61$&$0.10$&$78.99\pm 1.07$& $11.31\pm 0.91$& $11.21\pm 0.61$&
 $15.99\pm 0.55$& $4.64\pm 0.59$\\ 
OGLE 1999 BUL 32&$0152420101$&$0.05$&$0.02$&$0.05$&$23.88\pm 0.46$& $3.71\pm 0.38$& $2.92\pm 0.22$&
 $4.71\pm 0.22$& $0.70\pm 0.20$\\ 
MACHO 96 BLG 5&$0305970101$&$0.09$&$0.07$&$0.10$&$32.85\pm 0.31$& $5.92\pm 0.64$& $3.76\pm 0.20$&
 $6.15\pm 0.24$& $1.26\pm 0.20$\\ 
LH VLA 2&$0554121301$&$0.01$&$0.11$&$0.07$&$84.08\pm 1.03$& $12.89\pm 1.95$& $9.60\pm 0.77$&
 $15.98\pm 0.92$& $1.77\pm 0.55$\\ 
RXJ2328.8+1453&$0502430301$&$0.02$&$0.06$&$0.13$&$71.93\pm 0.79$& $7.52\pm 0.63$& $10.48\pm 0.43$&
 $13.83\pm 0.39$& $2.93\pm 0.33$\\ 
CDFS&$0555780101$&$0.01$&$0.17$&$0.10$&$75.15\pm 1.10$& $9.26\pm 0.86$& $10.97\pm 0.60$&
 $14.48\pm 0.55$& $2.19\pm 0.42$\\ 
CDFS&$0555780201$&$0.01$&$0.15$&$0.12$&$78.29\pm 0.75$& $13.70\pm 1.28$& $9.61\pm 0.58$&
 $15.29\pm 0.62$& $3.48\pm 0.40$\\ 
CDFS&$0555780301$&$0.02$&$0.14$&$0.14$&$77.36\pm 0.70$& $11.92\pm 1.23$& $9.82\pm 0.58$&
 $15.27\pm 0.60$& $3.25\pm 0.36$\\ 
CDFS&$0555780501$&$0.02$&$0.12$&$0.18$&$80.90\pm 0.67$& $11.07\pm 1.10$& $10.71\pm 0.55$&
 $15.89\pm 0.54$& $4.30\pm 0.35$\\ 
CDFS&$0555780601$&$0.02$&$0.11$&$0.14$&$83.02\pm 0.73$& $13.74\pm 1.35$& $10.34\pm 0.60$&
 $15.76\pm 0.65$& $3.14\pm 0.39$\\ 
CDFS&$0555780701$&$0.02$&$0.26$&$0.17$&$81.31\pm 0.75$& $11.71\pm 0.97$& $10.51\pm 0.56$&
 $16.36\pm 0.51$& $4.01\pm 0.35$\\ 
CDFS&$0555780801$&$0.02$&$0.13$&$0.14$&$87.24\pm 0.83$& $11.75\pm 1.27$& $10.78\pm 0.67$&
 $16.81\pm 0.65$& $3.76\pm 0.40$\\ 
CDFS&$0555780901$&$0.01$&$0.12$&$0.13$&$81.84\pm 0.81$& $12.51\pm 1.28$& $11.22\pm 0.66$&
 $16.49\pm 0.63$& $3.89\pm 0.41$\\ 
CDFS&$0555781001$&$0.02$&$0.16$&$0.17$&$84.51\pm 0.76$& $11.56\pm 1.12$& $10.63\pm 0.67$&
 $16.12\pm 0.59$& $3.74\pm 0.37$\\ 
CDFS&$0555782301$&$0.02$&$0.12$&$0.15$&$85.00\pm 0.77$& $11.66\pm 1.12$& $11.82\pm 0.60$&
 $16.59\pm 0.58$& $3.09\pm 0.38$\\ 

\hline
\end{tabular}
}
\caption{\label{tab:astrophysical} This table shows the best-fit values of the normalisations of spectral emission in the null hypothesis modell $\bf{m}_{\bf{0}}$. The mean energy $E_{\text{line}}$ of the spectral lines with Gaussian shape as well as their physical origins are presented in the first two rows. The observation identification (obsid), the radial distance $\phi$ and the weighting $w_{\text{NFW}}$ of the additional line are also listed.}
\end{table}

\end{document}